\documentclass[%
 reprint,
 amsmath, amssymb,
 aps, onecolumn, floatfix, superscriptaddress
]{revtex4-2}

\usepackage{graphicx}
\usepackage{dcolumn}
\usepackage{bm}
\usepackage{color}
\usepackage{multirow}
\usepackage{url}
\usepackage{float}

\begin{document}

\title{Are EEG functional networks really describing the brain? A comparison with other information-processing complex systems}

\author{Sofia Gil-Rodrigo}
\affiliation{Center for Brain and Cognition, Department and School of Engineering, Universitat Pompeu Fabra, 08005 Barcelona, Spain }

\author{Ra\'ul L\'opez-Mart\'in}
\affiliation{Instituto de F\'isica Interdisciplinar y Sistemas Complejos IFISC (CSIC-UIB), Campus UIB, 07122 Palma de Mallorca, Spain}

\author{G\"orsev Yener}
\affiliation{Department of Neurology, Faculty of Medicine, Dokuz Eyl\"ul University, Izmir, T\"urkiye }
\affiliation{IBG: International Biomedicine and Genome Center, Izmir, T\"urkiye }

\author{Jan R. Wiersema}
\affiliation{Department of Experimental Clinical and Health Psychology, Ghent University, Ghent, Belgium}

\author{Bahar G\"untekin}
\affiliation{Neuroscience Research Center, Research Institute for Health Sciences and Technologies (SABITA), Istanbul Medipol University, Istanbul, T\"urkiye }
\affiliation{Department of Biophysics, School of Medicine, Istanbul Medipol University, Istanbul, T\"urkiye }

\author{Massimiliano Zanin}
\thanks{Corresponding author: massimiliano.zanin@gmail.com}
\affiliation{ Instituto de F\'isica Interdisciplinar y Sistemas Complejos IFISC (CSIC-UIB), Campus UIB, 07122 Palma de Mallorca, Spain}

\begin{abstract}
Functional networks representing human brain dynamics have become a standard tool in neuroscience, providing an accessible way of depicting the computation performed by the brain in healthy and pathological conditions. Yet, these networks share multiple characteristics with those representing other natural and man-made complex systems, leading to the question of whether they are actually capturing the uniqueness of the human brain. By resorting to a large set of data representing multiple financial, technological, social, and natural complex systems, and by relying on Deep Learning classification models, we show how they are highly similar. We specifically reach the conclusion that, under some general reconstruction methodological choices, it is as difficult to understand whether a network represents a human brain or a financial market, as to diagnose a major pathology. This suggests that functional networks are describing information processing mechanisms that are common across complex systems; but that are not currently defining the uniqueness of the human mind. We discuss the consequence of these findings for neuroscience and complexity science in general, and suggest future avenues for exploring this interesting topic.
\end{abstract}

\keywords{EEG $|$ Functional brain networks $|$ Complex systems $|$ Deep Learning}

\maketitle

One of the most promising ways of representing the flow of information in a complex system is provided by functional complex networks, and, not surprisingly, these have strongly impacted neuroscience in the last decade. Beyond illustrating the structures created by information processing in healthy and pathological conditions \cite{bullmore2009complex, park2013structural}, functional brain networks have also been found to share many topological characteristics with similar network representations of other complex systems. As already highlighted in Ref. \cite{vertes2011topological}, both brain and financial networks are non-random, small-world, modular and hierarchical. This, on the one hand, can be explained by the fact that both systems are the results of similar processes, i.e. investors communicate and compute information in a way akin to neurons \cite{cristelli2013complexity}. 
Yet, on the other hand, this may also be the result of the default approach of understanding the brain via metaphors, hence via similarities with other systems. Since early times, understanding the brain was fuelled by metaphors corresponding to technical and technological systems that were at that time available and thought to be understood \cite{kirkland2002high, martensen2004brain}. To illustrate, Aristotle thought of the brain as a blood cooling machine; Descartes saw the brain as a hydraulic system; further metaphors included the clock, electricity, the telegraph, and computers - including working memory, information processing, computational capabilities, and so on. More recently the brain is understood as a network, not surprisingly, as our societies are built on top of networked systems \cite{barney2004network}. In other words, the history of neuroscience has been the history of these metaphors. This metaphor-based approach has shown its value, but is also limited by construction, as it cannot give us more insights than what we know about the metaphor. We may be able to find useful similarities but it will not reveal the uniqueness (and complexity) of the system under study, in this case, of the human brain. 

The human brain is not like any other complex system; on the contrary, it is the only system we know of that displays authentic intelligence, is self-aware, and is even able to write, read, and understand a scientific paper. This fact leads to two major questions. Firstly, how much are functional brain networks really representative of the human brain dynamics, as opposed to the general dynamics of information-processing complex systems. Secondly, and consequence of this, how representative are these networks of alterations emerging in pathologies. In other words, while Aristotle could only leverage limited tools and had to use a hypothesis driven/top down approach, nowadays we do have other possibilities, and we specifically can apply strong bottom up/data driven approaches to detect unique features and characteristics not present in other complex systems.

Unfold such new approach in the context of brain functional networks requires facing two methodological problems. Firstly, an unbiased and non-parametric way of analysing and comparing functional networks is needed. To illustrate, observing no differences in a given topological metric only indicates that such specific property is similar, but differences affecting other topological aspects cannot be excluded. In other words, a complete analysis would require testing all possible (both known and yet to be proposed) topological metrics \cite{costa2007characterization}. Secondly, the result of reconstructing functional networks strongly depends on the parameter used throughout the process, for which in many cases there are no theoretical guidelines, and whose values are left to the practitioner's judgement \cite{papo2014functional}.

We here address these two problems by, on the one hand, using Deep Learning models to assess how similar two sets of functional networks are, i.e. whether or not there are structural elements supporting their differentiation. DL models are hypotheses-free by construction, as they build a tailored abstract representation of the data. On the other hand, we guide the reconstruction of functional brain networks through their capacity of discriminating between healthy subjects and patients suffering from Parkinson's Disease (PD). These networks will then be compared against the functional representations of other complex systems, here drawn from financial, technological, social, and natural ones. In other words, considered brain functional networks are not tailored to this problem, but are rather optimised according to a typical neuroscience question. We finally pose the question: are we really capturing the uniqueness of the human brain?

\section*{Results}

We start by reconstructing functional networks representing brain dynamics of control subjects and PD patients, from resting state EEG recordings (Materials and Methods, SI Appendix I.A). For each subject and group, short segments of length $l$ of EEG time series are extracted, and a functional network is reconstructed therefrom using four different synchronisation measures (Materials and Methods). We further consider two possibilities: unweighted networks, where links are binarised according to a threshold $\tau$; and weighted ones, in which links are deleted if their weight is below $\tau$, but whose weight is preserved if passing this filter. Two sets of $500$ functional networks (one set per group, for a total of $10^3$ networks) are then used to train a Graph Isomorphism Network (GIN) DL model \cite{xu2019gin} (Materials and Methods), which is then evaluated over two independent (but similarly constructed) sets of networks. The result, measured in terms of the median classification accuracy over $500$ independent realisations, indicates whether the two sets are structurally similar, or, on the contrary, present some unique and characteristic structures.

\begin{figure*}[!t]
\centering
\includegraphics[width=17.8cm]{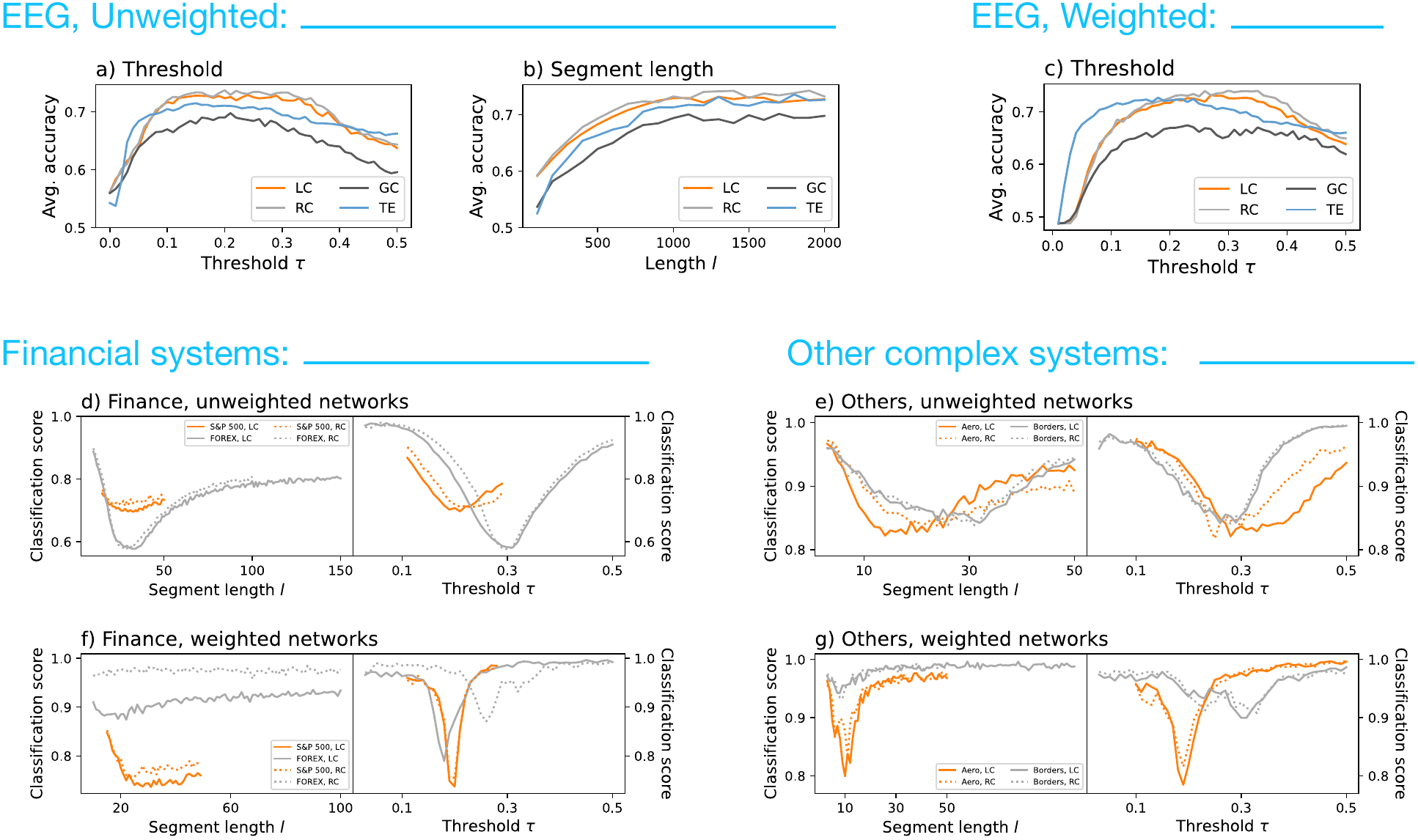}
\caption{ Optimising the reconstruction of functional networks. a)-b) Classification score between unweighted functional networks of control subjects and PD patients, as a function of the threshold $\tau$, the segment length $l$, and the synchronisation metric (line colours, see legends, and Materials and Methods for details). c) Classification results for EEG weighted functional networks, as a function of the threshold $\tau$ and the synchronisation metric ($l = 1400$). d)-g) Classification score between brain functional networks of control subjects, and functional networks of other complex systems, both unweighted (top panels) and weighted (bottom panels), as a function of the segment length $l$ and the threshold $\tau$. }
\label{fig:AllResults}
\end{figure*}

The evolution of the accuracy as a function of $\tau$, $l$, and the synchronisation measure, is reported in Fig. \ref{fig:AllResults} a) and b) for unweighted networks. As the objective is to use the best network representation to discriminate the two conditions, we set $\tau = 0.2$, $l = 1,400$, and Rank Correlation in subsequent analyses of unweighted networks. Note that filtering the data in specific frequency bands does not improve the discrimination (SI Appendix II.A). A similar analysis is performed for the weighted case, using the previously obtained segment length of $l = 1,400$, obtaining a maximum discrimination capability for $\tau = 0.3$ and Rank Correlation - see Fig. \ref{fig:AllResults} c).

Using these networks as a reference, we then move to the analysis of financial markets (SI Appendix I.B-C, Materials and Methods), with the aim of finding the functional networks representation that minimise the classification score when compared with the brain ones. In other words, we probe whether financial functional networks can be constructed to be as similar as possible to the dynamics observed in the brain. Panels d) and f) of Fig. \ref{fig:AllResults} report the evolution of the classification score for respectively unweighted and weighted networks, as a function of the threshold, segment lengths and synchronisation metric. Very low classification scores can be obtained, especially in the case of the unweighted versions of the networks, even below $0.6$. Similar analyses are reported in panels e) and g) for other complex systems (Materials and Methods, SI Appendix I.D-F). A synthesis of the best parameter sets is included in SI Appendix IV.D.

\begin{figure}[!tb]
\centering
\includegraphics[width=0.99\linewidth]{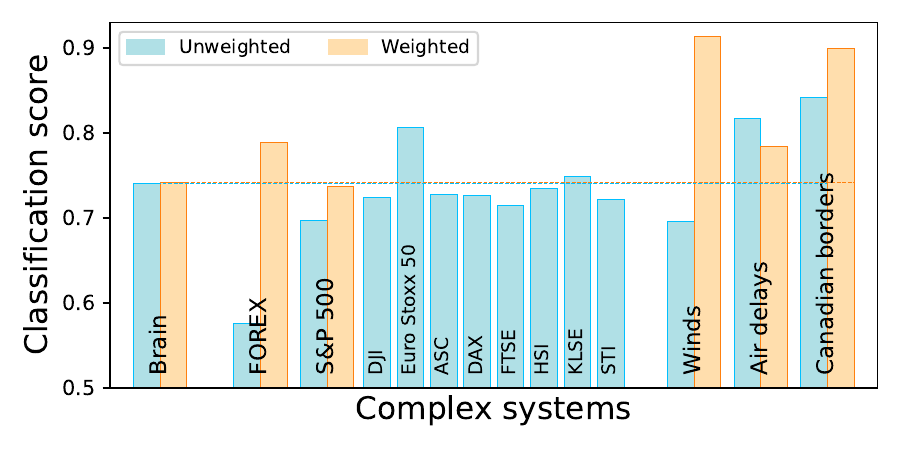}
\caption{ Synthesis of the best classification scores obtained across different complex systems. Blue and orange bars respectively correspond to results for unweighted and weighted network classifications. The two leftmost bars correspond to the classification between control subjects and PD patients, as in Fig. \ref{fig:AllResults} b) and c); for convenience, these reference levels are marked with two horizontal lines. }
\label{fig:Synthesis}
\end{figure}

The synthesis of these results, see Fig. \ref{fig:Synthesis}, highlights a key point. The similarity of functional networks of multiple complex systems to those representing brain activity of healthy individuals is on average the same as the similarity between the latter ones and those of PD patients. Most notably, the best case can be found for unweighted Forex networks, for which the classification score drops to $0.6$ - as opposed to the $0.75$ reported in Fig. \ref{fig:AllResults} a) and b). In other words: there are more differences between control subjects and Parkinson's patients, than between the same control subjects and the Forex market. This is not a side effect of using a DL classification model; an analysis of the topological metrics of both sets of networks confirms their similarity (SI Appendix V.C), with structures characterised by a core-periphery configuration, and by a high frequency of $3$- and $4$-nodes complete motifs (SI Appendix V.D). The reduced differences across complex systems is also not due to a lack in sensitivity of the DL model (SI Appendix VI).
The similarity between EEG and Forex networks can additionally be increased by performing a downsampling of the time series of the latter (SI Appendix VII), suggesting the presence of specific time scales in which similar computations may be performed by both systems. Note that a similar processing does not yield any benefit in the case of winds' networks (SI Appendix VII); and is further pointless in the other data sets, due to their lower time resolution.

\begin{figure}[!tb]
\centering
\includegraphics[width=0.99\linewidth]{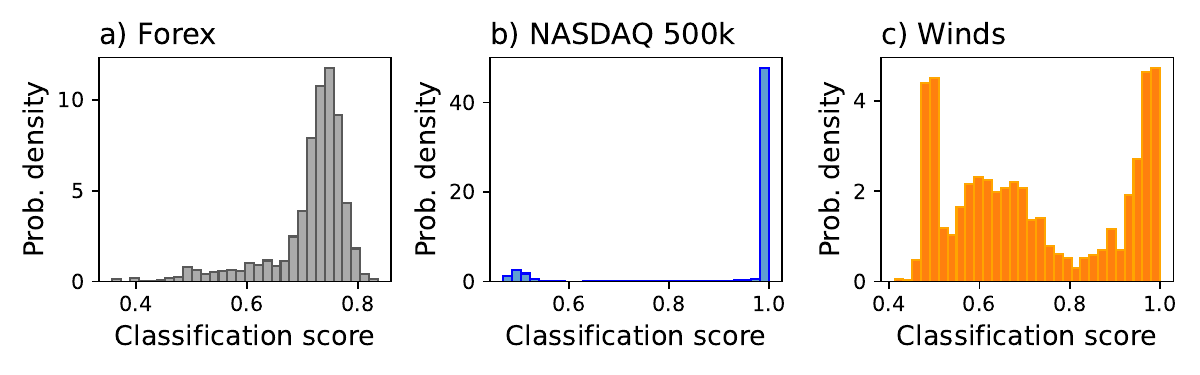}
\caption{ Analysis of extended data sets. The three panels report the probability density function of the classification score obtained when considered extended data sets for a) Forex, b) NASDAQ 500k, and c) winds data. See SI Appendix I for complete data set descriptions. }
\label{fig:Combinations}
\end{figure}

The previous results have been obtained by carefully tuning the parameters of the reconstruction process (i.e. $\tau$ and $l$) to yield networks as similar as possible to those representing brain dynamics. In order to explore how generalisable these results are, we firstly analyse networks representing other financial markets when these are reconstructed using the same $\tau$ and $l$ as optimised for the S\&P 500 data set. Results (see from DJI to STI in Fig. \ref{fig:Synthesis}, and SI Appendix IV.A) suggest no major increase in identifiability. In other words, all markets here considered have a similar structure that is resilient to the reconstruction procedure, and that is independent from main market characteristics (see SI Appendix V.A). We further analyse three data sets (Forex, NASDAQ 500k, and winds' data) for which we have access to a larger set of time series. Specifically, instead of selecting $30$ nodes according to some criteria, we sampled a large number of sets of $30$ randomly-chosen time series, and reconstructed the corresponding functional networks according to the optimised parameters used in Fig. \ref{fig:Synthesis}. Results reported in Fig. \ref{fig:Combinations} indicates that Forex and winds' data maintain a similar identifiability, with the latter case being weakly correlated to the spatial location of sensors (see SI Appendix V.B). On the other hand, NASDAQ 500k networks becomes highly identifiable; this is probably due to the inclusion of stocks of low market capitalisation, which may have a highly heterogeneous dynamics.

\section*{Discussion and conclusions}

Functional networks representing brain dynamics, and the dynamics of other complex systems, in many cases share similar structures. These similarities are highlighted by the difficulties encountered by state-of-the-art DL models to discriminate them, suggesting a low identifiability \cite{zanin2023identifiability}; and are further confirmed by comparable topological metrics, as previously highlighted in other works \cite{vertes2011topological}. Differences are in many cases, and especially in the case of financial networks, of the same magnitude as the differences between functional networks of control subjects and patients suffering from PD. In other words: it is as complex to discriminate between the two conditions, as to recognise whether one of these networks actually represents a human brain or a financial market. 

This surprising result generates many interesting research questions. First of all, where does this similarity come from? Is it a consequence of the way of reconstructing functional networks, or even of the concept of functional network itself? Or is rather a consequence of representing structures that are universal in information-processing systems? The fact that the maximum similarity is obtained for Forex networks, i.e. possibly the most complex and interconnected financial market, and is minimal in the Canadian borders data set, where a computation between travellers is actually not expected to exist, supports (but does not causally prove) the latter option. The relatively low identifiability of PD networks also supports this idea: while patients have an impaired cognition, this does not mean that no computation at all is performed by their brains; differences may therefore only stem from how altered such computation is. In other words, functional networks may only be the last step in the use of metaphors in neuroscience.

As a second question, given that brains and financial markets are very difficult to distinguish, are these functional networks really representing the uniqueness of the human brain? While the answer seems to be negative, this does not diminish the usefulness of functional networks in a neuroscience context. As widely shown in the literature, these can be used to elucidate information flows in the brain, both at rest and during cognitive tasks; and can further be used to detect pathological conditions. Yet, results here reported suggest (but, yet again, do not prove) that the resulting networks are not describing the uniqueness of the human brain, but rather common features of information-processing systems.

Thirdly, if the previous conjecture is true, a more complex (and even philosophical in nature) question remains: can functional networks be reconstructed, aimed at representing what is unique to the human brain? I.e., representing not only computation, but even our unique self-awareness and the theory of mind \cite{gallagher2003functional, frith2005theory, wellman2014making, nijhof2018brain}?

While the present work relies on state-of-the-art methods, especially regarding the assessment of the similarities between networks, it is clearly not free of limitations. On one hand, only one set of control subjects and one set of patients have been considered. Generalisability could be confirmed by using additional EEG recordings; and the inclusion of different brain pathologies of different degrees of severity, and of neurodivergent groups such as autism or attention deficit hyperactivity disorder, may shed light on when (if at all) the computation capability is reduced. The considered EEG data set comprises recording for only $30$ electrodes, and it may be suspected that this coarse-graining of brain dynamics may mask important local structures. In other words, low classification scores may be the result of a ``curse of dimensionality''. Yet, this would not be compatible with the high scores observed for air delays and Canadian borders' data sets. Time scales have shown to be an important element (see SI Appendix VII); data for complex systems of faster evolution, or even higher resolution data for those here considered, may improve our understanding of the previously posed questions. Finally, while GINs are the current state-of-the-art in DL models for classifying groups of networks, machine learning and artificial intelligence are fast evolving fields; better and more efficient solutions may appear in the near future. In order to support the scientific community in this last point, we include with this work a set of networks representing the worst obtained classification between control subjects and Forex, which can be used as benchmark in future studies (see SI Appendix VIII).
    
In conclusion, in this work we have leveraged DL models to highlight the similarities between functional networks representing human brain dynamics, and the dynamics of other complex systems. Presented results lead to an essential question: are functional networks really capturing the essence of human brain dynamics, or only patterns common to all information-processing systems? Furthermore, in the latter case, can we rely on functional networks to probe the dynamics of the human brain, with the hope of obtaining diagnostic and therapeutic knowledge? In other words: how can findings about the uniqueness of human brain impact our understanding (and hence diagnostics, support, and therapeutics) of neurodivergence and neuropathology? While the analyses here presented do not yield a conclusive answer, they surely highlight the necessity of delving deeper in our understanding of this essential tool.

\section*{Acknowledgments}
This project has received funding from the European Research Council (ERC) under the European Union's Horizon 2020 research and innovation programme (grant agreement No 851255). This work was partially supported by Grant CNS2023-144775 funded by MICIU/AEI/10.13039/501100011033 by ``European Union NextGenerationEU/PRTR'', and by the Mar\'ia de Maeztu project CEX2021-001164-M funded by the MICIU/AEI/10.13039/501100011033 and FEDER, EU.

\newpage
\clearpage

\section*{Materials and Methods}

\subsection*{Data sets}
We consider a large number of data sets representing the dynamics of human brain in normal and pathological (Parkinson's Disease) condition; and of other complex systems, spanning financial, technological, and social systems. These include:
\begin{itemize}
    \item Financial markets: Forex, i.e. the market for the trading of currencies; stocks composing the US S\&P 500 and DJI indices; stocks composing the European Euro Stoxx 50 index; companies classified as aggressive small caps (ASM) in the US NYSE stock exchange; and companies composing several national indices, including the German DAX, UK FTSE, Hong Kong's HSI, Malaysian's KLSE, and the Singaporean STI.
    \item Wind speed information recorded throughout the main European airports.
    \item Information about delays in major European airports.
    \item Daily number of foreign visitors entering Canada.
\end{itemize}

A full list of them, and of the elements thereof composed, is provided in SI Appendix I. Additionally, a description of the pre-processing carried out in each case is reported in SI Appendix II.

\subsection*{Functional network reconstruction}
Given a set of $30$ time series (as this is the number of available EEG electrodes), segments of length $l$ are extracted at random from them, and the weight of links between pairs of them is obtained using four standard metrics: linear correlation (LC), rank correlation (RC), Granger Causality (GC) \cite{granger1969investigating}, and Transfer Entropy (TE) \cite{schreiber2000measuring, vicente2011transfer} - see SI Appendix III for definitions. In the case of GC and TE, $p$-values are rescaled to yield values similar to those of LC and RC. Finally, two variants of networks are obtained: unweighted ones, in which all links with weight below a threshold $\tau$ are discarded, and the weight of surviving links is set to one; and weighted ones, in which the weight of links greater than $\tau$ is preserved - see SI Appendix III.

\subsection*{Deep Learning model and classification}
The Deep Learning model used for the network classification is the Graph Isomorphism Neural Network (GIN)~\cite{xu2019gin}, which is as powerful as the Weisfeiler-Lehman test regarding graph classification, enabling it to capture the concept of isomorphism~\cite{xu2019gin}. This makes it the most suitable tool for classifying networks based on their topology without introducing biases derived from focusing on specific arbitrary topological metrics. The GIN model was implemented using the Python \textit{Pytorch} library~\cite{pytorch} and its structure consists of three convolutional layers of dimension $(h,h)$ and two fully connected layers of size $3h$, where $h=32$.

The classification process began with the random selection of $500$ networks from each of the two classes. Using a batch size of $64$ networks and running for $5,000$ epochs, we performed $500$ classifications for each reconstruction parameter value, averaging the results to generate the curves shown in Fig.~\ref{fig:AllResults}. 

Initially, we classified control brain networks against PDs brain networks (see Fig.~\ref{fig:AllResults} (a-c)). For subsequent classifications of control brain networks against other complex systems, we maintained the optimal $l$ and $\tau$ that provided the best accuracy in distinguishing control subjects from PD. This approach assumes that the most effective representation in differentiating control and Parkinson's brain networks is also the best representation of a healthy brain. We then swept these parameters for the networks of the other class to obtain the remaining curves observed in Fig.~\ref{fig:AllResults}. Further details are reported case by case in SI Appendix IV, and a synthesis of the best parameters can be found in SI Appendix IV.D.


\newpage
\clearpage

\setcounter{figure}{0}

\section{Data sources}
\label{sec:data_sources}

\subsection{EEG data}

A total of 105 subjects, comprising both eyes-open and eyes-closed EEG recordings, were included in the study, divided between 66 healthy elderly controls and 39 Parkinson's disease patients with dementia (PD). All resting state EEG recordings were performed at Istanbul Medipol University, with subjects evaluated by an experienced neurologist. The Movement Disorder Society (MDS) Level 1 criteria \cite{dubois2007diagnostic,emre2007clinical} were used for the probable PD diagnosis, made when all the following five criteria were met: (i) PD diagnosis according to United  Kingdom Parkinson's Disease Society Brain Bank Criteria \cite{hughes1992accuracy}; (ii) development of PD prior to the onset of dementia; (iii) PD associated with a decreased global cognitive score that was defined as a score of $\leq 24$ on the MMSE \cite{folstein1975mini}; (iv) cognitive impairment that causes dysfunction in daily life; and (v) cognitive impairments found in more than one domain. All patients with PD were evaluated $60$-$90$ min after their morning dose of levodopa for the EEG recordings. 
Healthy elderly subjects were included when no neurological abnormality or no global cognitive impairment (Mini-mental State Examination (MMSE) score $\geq 27$), no history or presence of any psychiatric abnormalities, and no history of drug and alcohol abuse were determined.
EEG of patients and healthy controls were recorded for two different projects. The ethical committee of Istanbul Medipol University (No: 10840098-51, No:10840098-604) approved the studies; informed consent was obtained from all participants or caregivers.

A 32-channel EEG was recorded using a Brain-Amp 32-channel DC system machine with Ag/AgCl electrodes. Two earlobe electrodes were used as reference electrodes. EEG was recorded with $0.01$-$250$Hz band limits and a $500$Hz sampling rate. Four minutes of eyes closed and four minutes of eyes open EEG was recorded from all subjects. EEG was recorded in a dimly lit isolated room, and a video camera controlled the subjects during the recordings.

\subsection{Stock markets' data}

Four different data sets representing the dynamics of stock markets have been considered, specifically:

\begin{itemize}
    \item The $30$ largest companies, in terms of market capitalisation, composing the {\bf S\&P 500} (Standard and Poor's 500) index. Analysed data correspond to the time period from February $1^{st}$ $2013$ to February $2^{nd}$ $2024$, for a total of $2,790$ data points. The full list is reported in Tab. \ref{tab:sp500}.
    \item The $30$ largest companies, in terms of market capitalisation, composing the German {\bf DAX} (Deutscher Aktienindex) index. Analysed data correspond to the time period from September $30^{th}$ $2022$ to April $30^{th}$ $2024$, for a total of $403$ data points. The full list is reported in Tab. \ref{tab:dax}.
    \item The $30$ largest companies classified as {\bf aggressive small caps (ASM)} in the NYSE stock exchange. These are defined as companies of small capitalisation (below \$2 billion USD) and with an increase in the EPS (earnings per share) in the last year greater than $25\%$. Analysed data correspond to the time period from December $16^{th}$ $2020$ to May $1^{st}$ $2024$, for a total of $847$ data points. The full list is reported in Tab. \ref{tab:asc}.
    \item The $30$ largest companies, in terms of market capitalisation, composing the UK {\bf FTSE 100} (Financial Times Stock Exchange 100 Index) index. Analysed data correspond to the time period from July $18^{th}$ $2022$ to May $1^{st}$ $2024$, for a total of $451$ data points. The full list is reported in Tab. \ref{tab:ftse}.
    \item The $30$ largest companies, in terms of market capitalisation, composing the European {\bf Euro Stoxx 50} index. Analysed data correspond to the time period from September $3^{rd}$ $2001$ to February $2^{nd}$ $2024$, for a total of $5,710$ data points. The full list is reported in Tab. \ref{tab:ftse}.
    \item The $30$ largest companies, in terms of market capitalisation, composing the {\bf Dow Jones Industrial Average (DJI)} index. Analysed data correspond to the time period from March $20^{th}$ $2019$ to February $2^{nd}$ $2024$, for a total of $1,227$ data points. The full list is reported in Tab. \ref{tab:dji}.
    \item The $30$ largest companies, in terms of market capitalisation, composing the Hong Kong's {\bf Hang Seng (HSI)} index. Analysed data correspond to the time period from December $23^{rd}$ $2020$ to April $30^{th}$ $2024$, for a total of $882$ data points. The full list is reported in Tab. \ref{tab:hsi}.
    \item The $30$ largest companies, in terms of market capitalisation, composing the Malaysian's {\bf FTSE Bursa Malaysia (KLSE)} index. Analysed data correspond to the time period from October $26^{th}$ $2020$ to April $30^{th}$ $2024$, for a total of $852$ data points. The full list is reported in Tab. \ref{tab:klse}.
    \item The $30$ companies composing the {\bf Straight Times (STI)} index, representing the largest companies in the Singapore Exchange market. Analysed data correspond to the time period from July $14^{th}$ $2022$ to April $30^{th}$ $2024$, for a total of $450$ data points. The full list is reported in Tab. \ref{tab:sti}.
\end{itemize}

In all cases, data have been obtained from Yahoo! Finance and are freely available at \url{https://finance.yahoo.com}. Market capitalisation corresponds to the last day in each data set. Each stock is represented by a time series, measuring the adjusted (for dividends, stock splits, and new stock offerings) daily closing price. Note that the initial date of each data set is conditioned by the first day in which all the companies composing it were quoted, hence the great variability in the number of available points.

\begin{table*}[!tb]
\caption{\label{tab:sp500} List of the $30$ considered stocks part of the {\bf S\&P 500} index, sorted by alphabetic order; for each stock, the corresponding code and full names are reported. }
{\small
\begin{tabular}{|p{1.5cm} p{4.0cm}|p{1.5cm} p{4.0cm}|p{1.5cm} p{4.0cm}|}
    \hline
    AAPL & Apple Inc & ABBV & AbbVie Inc & ADBE & Adobe Inc \\ \hline 
    AMD & AMD & AMZN & Amazon.com Inc & AVGO & Broadcom Inc \\ \hline
    BRK-B & Berkshire Hathaway Inc Class B & COST & Costco Wholesale Corporation & CRM & Salesforce Inc \\ \hline
    CVX & Chevron Corp & GOOG & Alphabet Inc Class C & GOOGL & Alphabet Inc Class A \\ \hline
    HD & Home Depot Inc & JNJ & Johnson \& Johnson & JPM & JPMorgan Chase \& Co \\ \hline 
    KO & Coca-Cola Co & LLY & Eli Lilly And Co & MA & Mastercard Inc \\ \hline
    META & Meta Platforms Inc & MRK & Merck \& Co Inc & MSFT & Microsoft Corp \\ \hline
    NFLX & Netflix Inc & NVDA & NVIDIA Corp & PEP & PepsiCo Inc \\ \hline
    PG & Procter \& Gamble Co & TSLA & Tesla Inc & UNH & UnitedHealth Group Inc \\ \hline
    V & Visa Inc & WMT & Walmart Inc & XOM & Exxon Mobil Corp \\ \hline 
\end{tabular}
}
\end{table*}

\begin{table*}[!tb]
\caption{\label{tab:dax} List of the $30$ considered stocks part of the {\bf DAX} index, sorted by alphabetic order; for each stock, the corresponding code and full names are reported. }
{\small
\begin{tabular}{|p{1.8cm} p{5.0cm}|p{1.8cm} p{5.0cm}|}
    \hline
1COV.DE & Covestro AG & AIR.DE & Airbus SE \\ \hline 
ADS.DE & Adidas AG & ALV.DE & Allianz SE \\ \hline 
BAS.DE & BASF SE & BAYN.DE & Bayer AG \\ \hline 
BEI.DE & Beiersdorf AG & BMW.DE & Bayerische Motoren Werke AG \\ \hline 
CON.DE & Continental AG & DB1.DE & Deutsche B\"orse AG \\ \hline 
DBK.DE & Deutsche Bank AG & DHL.DE & Deutsche Post AG \\ \hline 
DTE.DE & Deutsche Telekom AG & DTG.DE & Daimler Truck Holding AG \\ \hline 
ENR.DE & Siemens Energy AG & EOAN.DE & E.ON SE \\ \hline 
FRE.DE & Fresenius SE & HEI.DE & Heidelberg Materials AG \\ \hline 
HNR1.DE & Hannover R\"uck SE & IFX.DE & Infineon Technologies AG \\ \hline 
MRK.DE & Merck KGaA & MTX.DE & MTU Aero Engines AG \\ \hline 
P911.DE & Dr Ing hc F Porsche AG & RWE.DE & RWE AG \\ \hline 
SHL.DE & Siemens Healthineers AG & SIE.DE & Siemens AG \\ \hline 
SY1.DE & Symrise AG & VNA.DE & Vonovia SE \\ \hline 
VOW3.DE & Volkswagen AG & ZAL.DE & Zalando SE \\ \hline 
\end{tabular}
}
\end{table*}

\begin{table*}[!tb]
\caption{\label{tab:asc} List of the $30$ considered stocks part of the {\bf Aggressive small caps} index, sorted by alphabetic order; for each stock, the corresponding code and full names are reported. }
{\small
\begin{tabular}{|p{1.8cm} p{5.5cm}|p{1.8cm} p{5.5cm}|}
    \hline
    DHT & DHT Holdings Inc & ADUS & Addus Homecare Corporation \\ \hline
    AGIO & Agios Pharmaceuticals Inc & AMBA & Ambarella Inc \\ \hline
    BRP & BRP Group Inc & CAKE & Cheesecake Factory Inc \\ \hline
    CEPU & Central Puerto ADR & CRTO & Criteo SA \\ \hline
    EPAC & Enerpac Tool Group Corp & GNL & Global Net Lease Inc \\ \hline
    GSBD & Goldman Sachs BDC Inc & HEES & H\&E Equipment Services, Inc. \\ \hline
    JBLU & JetBlue Airways Corporation & JELD & Jeld-Wen Holding Inc \\ \hline
    LPG & Dorian LPG Ltd & MRCY & Mercury Systems Inc \\ \hline
    MYGN & Myriad Genetics, Inc. & NGL-PB & NGL Energy Partners LP \\ \hline
    NGL-PC & NGL Energy Partners LP & PARR & Par Pacific Holdings Inc \\ \hline
    PEB & Pebblebrook Hotel Trust & PLMR & Palomar Holdings Inc \\ \hline
    SAH & Sonic Automotive Inc & TBBK & Bancorp Inc \\ \hline
    TDS & Telephone and Data Systems Inc & TNK & Teekay Tankers Ltd \\ \hline
    TRMK & Trustmark Corp & UPST & Upstart Holdings Inc \\ \hline
    USLM & United States Lime \& Minerals Inc & YY & JOYY Inc \\ \hline
    
\end{tabular}
}
\end{table*}

\begin{table*}[!tb]
\caption{\label{tab:ftse} List of the $30$ considered stocks part of the {\bf FTSE 100} index, sorted by alphabetic order; for each stock, the corresponding code and full names are reported. }
{\small
\begin{tabular}{|p{1.8cm} p{6.0cm}|p{1.8cm} p{5.5cm}|}
    \hline
CCH.L & Coca-Cola HBC AG & AAF.L & Airtel Africa plc \\ \hline
AHT.L & Ashtead Group plc & ANTO.L & Antofagasta plc \\ \hline
AUTO.L & Auto Trader Group plc & BA.L & BAE Systems plc \\ \hline
BATS.L & British American Tobacco plc & CNA.L & Centrica plc \\ \hline
CPG.L & Compass Group plc & DPLM.L & Diploma plc \\ \hline
ENT.L & Entain plc  & EXPN.L & Experian plc \\ \hline
EZJ.L & easyJet plc & HLN.L & Haleon plc \\ \hline
ICG.L & Intermediate Capital Group plc & MNG.L & M\&G plc \\ \hline
PRU.L & Prudential plc & PSN.L & Persimmon plc \\ \hline
RMV.L & Rightmove plc & RR.L & Rolls-Royce Holdings plc \\ \hline
RTO.L & Rentokil Initial plc & SDR.L & Schroders plc \\ \hline
SHEL.L & Shell plc & SMIN.L & Smiths Group plc \\ \hline
SMT.L & Scottish Mortgage Investment Trust plc & SPX.L & Spirax-Sarco Engineering plc \\ \hline
SSE.L & SSE plc & STJ.L & St. James's Place plc \\ \hline
TSCO.L & Tesco plc & VOD.L & Vodafone Group plc \\ \hline
\end{tabular}
}
\end{table*}

\begin{table*}[!tb]
\caption{\label{tab:es500} List of the $30$ considered stocks part of the {\bf Euro Stoxx 50} index, sorted by alphabetic order; for each stock, the corresponding code and full names are reported. }
{\small
\begin{tabular}{|p{1.5cm} p{4.0cm}|p{1.5cm} p{4.0cm}|p{1.5cm} p{4.0cm}|}
    \hline
ABI.BR & Anheuser-Busch InBev SA/NV & AI.PA & L'Air Liquide S.A. & AIR.PA & Airbus SE \\ \hline
ALV.DE & Allianz SE & ASML.AS & ASML Holding N.V. & BAS.DE & BASF SE \\ \hline
BAYN.DE & Bayer Aktiengesellschaft & BBVA.MC & Banco Bilbao Vizcaya Argentaria, S.A. & BMW.DE & Bayerische Motoren Werke Aktiengesellschaft \\ \hline
BN.PA & Danone S.A. & BNP.PA & BNP Paribas SA & CA.PA & Carrefour SA \\ \hline
CS.PA & AXA SA & DB1.DE & Deutsche B\"orse AG & DBK.DE & Deutsche Bank Aktiengesellschaft \\ \hline
DG.PA & Vinci SA & DHL.DE & Deutsche Post AG & DTE.DE & Deutsche Telekom AG \\ \hline
EL.PA & EssilorLuxottica SA & ENEL.MI & Enel SpA & ENGI.PA & Engie SA \\ \hline
ENI.MI & Eni S.p.A. & EOAN.DE & E.ON SE & FRE.DE & Fresenius SE \& Co. KGaA \\ \hline
G.MI & Assicurazioni Generali S.p.A. & GLE.PA & Soci\'et\'e G\'en\'erale SA & BE.MC & Iberdrola, S.A. \\ \hline
INGA.AS & ING Groep N.V. & ISP.MI & Intesa Sanpaolo S.p.A. & RI.PA & Pernod Ricard SA \\ \hline

\end{tabular}
}
\end{table*}

\begin{table*}[!tb]
\caption{\label{tab:dji} List of the $30$ considered stocks part of the {\bf Down Jones Industrial Average (DJI)} index, sorted by alphabetic order; for each stock, the corresponding code and full names are reported. }
{\small
\begin{tabular}{|p{1.5cm} p{4.0cm}|p{1.5cm} p{4.0cm}|p{1.5cm} p{4.0cm}|}
    \hline
AAPL & Apple Inc. & AMGN & Amgen Inc. & AXP & American Express Company \\ \hline
BA & The Boeing Company & CAT & Caterpillar Inc. & CRM & Salesforce, Inc. \\ \hline
CSCO & Cisco Systems, Inc. & CVX & Chevron Corporation & DIS & The Walt Disney Company \\ \hline
DOW & Dow Inc. & GS & The Goldman Sachs Group, Inc. & HD & The Home Depot, Inc. \\ \hline
HON & Honeywell International Inc. & IBM & International Business Machines Corporation & INTC & Intel Corporation \\ \hline
JNJ & Johnson \& Johnson & JPM & JPMorgan Chase \& Co. & KO & The Coca-Cola Company \\ \hline
MCD & McDonald's Corporation & MMM & 3M Company & MRK & Merck \& Co., Inc. \\ \hline
MSFT & Microsoft Corporation & NKE & NIKE, Inc. & PG & The Procter \& Gamble Company \\ \hline
TRV & The Travelers Companies, Inc. & UNH & UnitedHealth Group Incorporated & V & Visa Inc. \\ \hline
VZ & Verizon Communications Inc. & WBA & Walgreens Boots Alliance, Inc. & WMT & Walmart Inc. \\ \hline

\end{tabular}
}
\end{table*}

\begin{table*}[!tb]
\caption{\label{tab:hsi} List of the $30$ considered stocks part of the {\bf Hang Seng Index (HSI)} index, sorted by alphabetic order; for each stock, the corresponding code and full names are reported. }
{\small
\begin{tabular}{|p{1.5cm} p{7.0cm}|p{1.5cm} p{7.0cm}|}
    \hline
0002.HK & CLP Holdings Limited & 0003.HK & The Hong Kong and China Gas Company Limited \\ \hline
0012.HK & Henderson Land Development Company Limited & 0017.HK & New World Development Company Limited \\ \hline
0027.HK & Galaxy Entertainment Group Limited & 0101.HK & Hang Lung Properties Limited \\ \hline
0241.HK & Alibaba Health Information Technology Limited & 0267.HK & CITIC Limited \\ \hline
0669.HK & Techtronic Industries Company Limited & 0883.HK & CNOOC Limited \\ \hline
0992.HK & Lenovo Group Limited & 1038.HK & CK Infrastructure Holdings Limited \\ \hline
1044.HK & Hengan International Group Company Limited & 1093.HK & CSPC Pharmaceutical Group Limited \\ \hline
1109.HK & China Resources Land Limited & 1209.HK & China Resources Mixc Lifestyle Services Limited \\ \hline
1398.HK & Industrial and Commercial Bank of China Limited & 1810.HK & Xiaomi Corporation \\ \hline
2020.HK & ANTA Sports Products Limited & 2269.HK & WuXi Biologics (Cayman) Inc. \\ \hline
2319.HK & China Mengniu Dairy Company Limited & 2331.HK & Li Ning Company Limited \\ \hline
2628.HK & China Life Insurance Company Limited & 2688.HK & ENN Energy Holdings Limited \\ \hline
3690.HK & Meituan & 6098.HK & Country Garden Services Holdings Company Limited \\ \hline
6690.HK & Haier Smart Home Co., Ltd. & 9618.HK & JD.com, Inc. \\ \hline
9633.HK & Nongfu Spring Co., Ltd. & 9988.HK & Alibaba Group Holding Limited \\ \hline

\end{tabular}
}
\end{table*}

\begin{table*}[!tb]
\caption{\label{tab:klse} List of the $30$ considered stocks part of the {\bf FTSE Bursa Malaysia (KLSE)} index, sorted by alphabetic order; for each stock, the corresponding code and full names are reported. }
{\small
\begin{tabular}{|p{1.5cm} p{7.0cm}|p{1.5cm} p{7.0cm}|}
    \hline

1015.KL & AMMB Holdings Berhad & 1023.KL & CIMB Group Holdings Berhad \\ \hline
1066.KL & RHB Bank Berhad & 1082.KL & Hong Leong Financial Group Berhad \\ \hline
1155.KL & Malayan Banking Berhad & 1295.KL & Public Bank Berhad \\ \hline
1961.KL & IOI Corporation Berhad & 2445.KL & Kuala Lumpur Kepong Berhad \\ \hline
3182.KL & Genting Berhad & 3816.KL & MISC Berhad \\ \hline
4065.KL & PPB Group Berhad & 4197.KL & Sime Darby Berhad \\ \hline
4677.KL & YTL Corporation Berhad & 4707.KL & Nestl\'e (Malaysia) Berhad \\ \hline
4715.KL & Genting Malaysia Berhad & 4863.KL & Telekom Malaysia Berhad \\ \hline
5183.KL & PETRONAS Chemicals Group Berhad & 5225.KL & IHH Healthcare Berhad \\ \hline
5285.KL & SD Guthrie Berhad & 5296.KL & Mr D.I.Y. Group (M) Berhad \\ \hline
5347.KL & Tenaga Nasional Berhad & 5681.KL & PETRONAS Dagangan Berhad \\ \hline
5819.KL & Hong Leong Bank Berhad & 6012.KL & Maxis Berhad \\ \hline
6033.KL & PETRONAS Gas Berhad & 6742.KL & YTL Power International Berhad \\ \hline
6888.KL & Axiata Group Berhad & 6947.KL & Celcomdigi Berhad \\ \hline
7084.KL & QL Resources Berhad & 8869.KL & Press Metal Aluminium Holdings Berhad \\ \hline   
\end{tabular}
}
\end{table*}

\begin{table*}[!tb]
\caption{\label{tab:sti} List of the $30$ considered stocks part of the {\bf Straits Times Index (STI)} index, sorted by alphabetic order; for each stock, the corresponding code and full names are reported. }
{\small
\begin{tabular}{|p{1.5cm} p{7.0cm}|p{1.5cm} p{7.0cm}|}
    \hline

9CI.SI & CapitaLand Investment Limited & AJBU.SI & Keppel DC REIT \\ \hline
BN4.SI & Keppel Ltd. & BS6.SI & Yangzijiang Shipbuilding (Holdings) Ltd. \\ \hline
BUOU.SI & Frasers Logistics \& Commercial Trust & C07.SI & Jardine Cycle \& Carriage Limited \\ \hline
C09.SI & City Developments Limited & C38U.SI & CapitaLand Integrated Commercial Trust \\ \hline
C52.SI & ComfortDelGro Corporation Limited & D01.SI & DFI Retail Group Holdings Limited \\ \hline
D05.SI & DBS Group Holdings Ltd & EMI.SI & Emperador Inc. \\ \hline
F34.SI & Wilmar International Limited & G13.SI & Genting Singapore Limited \\ \hline
H78.SI & Hongkong Land Holdings Limited & J36.SI & Jardine Matheson Holdings Limited \\ \hline
J69U.SI & Frasers Centrepoint Trust & M44U.SI & Mapletree Logistics Trust \\ \hline
ME8U.SI & Mapletree Industrial Trust & N2IU.SI & Mapletree Pan Asia Commercial Trust \\ \hline
O39.SI & Oversea-Chinese Banking Corporation Limited & S51.SI & Seatrium Limited \\ \hline
S58.SI & SATS Ltd. & S63.SI & Singapore Technologies Engineering Ltd \\ \hline
U14.SI & UOL Group Limited & U96.SI & Sembcorp Industries Ltd \\ \hline
V03.SI & Venture Corporation Limited & Y92.SI & Thai Beverage Public Company Limited \\ \hline
YF8.SI & Yangzijiang Financial Holding Ltd. & Z74.SI & Singapore Telecommunications Limited \\ \hline
\end{tabular}
}
\end{table*}

\subsection{Forex' data}

Forex stands for foreign exchange market, and is a global decentralised or over-the-counter market for the trading of currencies. It has some unique characteristics, including its large trading volume, its geographical dispersion, and its continuous operation (24 hours a day except for weekends). In this study we have considered two sets of $30$ main currency pairs, reported in Tab. \ref{tab:forex}. Data for each pair have been obtained from \url{HistData.com}, and represent time series of prices with a 1 minute time resolution. In all cases, analysed data correspond to the period January 1st 2011 to December 31st 2023.

\begin{table*}[!tb]
\caption{\label{tab:forex} List of the $30 + 30$ considered currency pairs of the {\bf Forex} market, sorted by alphabetic order. Names corresponding to each code are reported in the bottom part. }
{\small
\begin{tabular}{|p{2.2cm}|p{2.2cm}|p{2.2cm}|p{2.2cm}|p{2.2cm}|p{2.2cm}|}
    \hline
    AUD-CAD & AUD-CHF & AUD-JPY & AUD-USD & AUX-AUD & CAD-JPY \\ \hline
    CHF-JPY & EUR-AUD & EUR-CAD & EUR-CHF & EUR-GBP & EUR-JPY \\ \hline
    EUR-NZD & EUR-USD & GBP-CHF & GBP-JPY & GBP-USD & GRX-EUR \\ \hline
    NZD-CAD & NZD-JPY & NZD-USD & SGD-JPY & USD-CAD & USD-CHF \\ \hline
    USD-HKD & USD-JPY & USD-MXN & USD-NOK & USD-TRY & XAU-USD \\ \hline 

    \multicolumn{6}{l}{} \\ \hline
    AUD-NZD & BCO-USD & CAD-CHF & EUR-CZK & EUR-DKK & EUR-HUF \\ \hline
    EUR-NOK & EUR-PLN & EUR-SEK & EUR-TRY & FRX-EUR & GBP-AUD \\ \hline
    GBP-CAD & GBP-NZD & HKX-HKD & JPX-JPY & NSX-USD & NZD-CHF \\ \hline
    SPX-USD & UDX-USD & UKX-GBP & USD-CZK & USD-DKK & USD-HUF \\ \hline
    USD-PLN & USD-SEK & USD-SGD & USD-ZAR & WTI-USD & ZAR-JPY \\ \hline

    \multicolumn{6}{l}{} \\ \hline
    \multicolumn{2}{|l|}{AUD: Australian Dollar} & 
    \multicolumn{2}{|l|}{AUX: ASX 200} & 
    \multicolumn{2}{|l|}{BCO: Brent crude oil} \\ \hline
    \multicolumn{2}{|l|}{FRX: CAC 40} & 
    \multicolumn{2}{|l|}{CAD: Canadian Dollar} & 
    \multicolumn{2}{|l|}{CZK: Czech Koruna} \\ \hline
    \multicolumn{2}{|l|}{DKK: Danish Krone} & 
    \multicolumn{2}{|l|}{GRX: DAX 30} & 
    \multicolumn{2}{|l|}{EUR: Euro} \\ \hline
    \multicolumn{2}{|l|}{XAU: Gold} & 
    \multicolumn{2}{|l|}{HKX: Han Seng} & 
    \multicolumn{2}{|l|}{HKD: Hong Kong Dollar} \\ \hline
    \multicolumn{2}{|l|}{HUF: Hungarian Forint} & 
    \multicolumn{2}{|l|}{JPY: Japanese Yen} & 
    \multicolumn{2}{|l|}{MXN: Mexican Peso} \\ \hline
    \multicolumn{2}{|l|}{NZD: New Zealand Dollar} & 
    \multicolumn{2}{|l|}{NOK: Norwegian Krone} & 
    \multicolumn{2}{|l|}{PLN: Polish z\l oty} \\ \hline
    \multicolumn{2}{|l|}{GBP: Pound sterling} & 
    \multicolumn{2}{|l|}{SGD: Singapore Dollar} & 
    \multicolumn{2}{|l|}{ZAR: South African Rand} \\ \hline
    \multicolumn{2}{|l|}{SEK: Swedish Krona} & 
    \multicolumn{2}{|l|}{CHF: Swiss Franc } & 
    \multicolumn{2}{|l|}{TRY: Turkish lira} \\ \hline
    \multicolumn{2}{|l|}{USD: United States Dollar} & 
    \multicolumn{2}{|l|}{WTI: West Texas Intermediate} & \multicolumn{2}{|l|}{} \\ \hline
    
\end{tabular}
}
\end{table*}

\subsection{Weather (wind) data}

Weather information has been extracted from METARs (METeorological Aerodrome Reports), i.e. reports describing the meteorological conditions at the largest European airports. These have been collected in real-time from the website \url{allmetsat.com} using an in-house software scraper with a time resolution of $30$ minutes; and are composed of a text following the standard defined by the World Meteorological Organisation (WMO) \cite{WMOTechReg}. 

Data were available from approximately April 25$^{th}$ 2019, to April $1^{st}$ 2022, for a total of more than $51,000$ reports.
A custom-made parser has been developed to extract and store basic information from each report, including: the 4-letters ICAO code of the issuing airport, the date and time of validity, temperature and dew point, and wind speed. Only the latter meteorological variable has been used in this study.

Additionally, two sets of airports have been considered: the $30$ largest ones in Europe (see Tab. \ref{tab:EUAirports}), plus a larger set of smaller airports (see Tab. \ref{tab:EUAirports2}) that are used to create extended networks.

\begin{table*}[!tb]
\caption{\label{tab:EUAirports} List of the $30$ largest EU airports for which meteorological data have been extracted, including both their 4-letters ICAO code (i.e. their unique identifier in the original data set) and the full official name. }
{\small
\begin{tabular}{|p{8.0cm}|p{8.0cm}|}
    \hline

EGLL: Heathrow Airport & 
LFPG: Charles de Gaulle Airport \\ \hline
EHAM: Amsterdam Airport Schiphol & 
EDDF: Frankfurt am Main Airport \\ \hline
LEMD: Adolfo Su\'arez Madrid-Barajas Airport & 
LEBL: Josep Tarradellas Barcelona-El Prat Airport \\ \hline
EDDM: Munich Airport & 
EGKK: Gatwick Airport \\ \hline
LIRF: Leonardo da Vinci-Fiumicino Airport & 
LFPO: Orly Airport \\ \hline
EIDW: Dublin Airport & 
LSZH: Zurich Airport \\ \hline
EKCH: Copenhagen Airport & 
LEPA: Palma de Mallorca Airport \\ \hline
ENGM: Oslo Airport, Gardermoen & 
EGCC: Manchester Airport \\ \hline
EGSS: London Stansted Airport & 
LOWW: Vienna International Airport \\ \hline
ESSA: Stockholm Arlanda Airport & 
EBBR: Brussels Airport \\ \hline
LIMC: Malpensa Airport & 
EDDL: D\"usseldorf Airport \\ \hline
LEMG: M\'alaga Airport & 
EPWA: Warsaw Chopin Airport \\ \hline
LSGG: Geneva Airport & 
EDDH: Hamburg Airport \\ \hline
LKPR: V\'aclav Havel Airport Prague & 
EGGW: Luton Airport \\ \hline
LHBP: Budapest Ferenc Liszt International Airport & 
EGPH: Edinburgh Airport \\ \hline
\end{tabular}
}
\end{table*}

\begin{table*}[!tb]
\caption{\label{tab:EUAirports2} List of the additional EU airports used to reconstruct extended wind networks. As in Tab. \ref{tab:EUAirports}, it includes the 4-letters ICAO code (i.e. their unique identifier in the original data set) and the full official name. }
{\small
\begin{tabular}{|p{8.0cm}|p{8.0cm}|}
    \hline

EBAW: Antwerp International Airport & 
EBCI: Brussels South Charleroi Airport \\ \hline
EBLG: Liege Airport & 
EBOS: Ostend-Bruges International Airport \\ \hline
EDDB: Berlin Brandenburg Airport & 
EDDC: Dresden Airport \\ \hline
EDDG: M\"unster Osnabr\"uck Airport & 
EDDK: Cologne Bonn Airport \\ \hline
EDDN: Nuremberg Airport & 
EDDP: Leipzig/Halle Airport \\ \hline
EDDS: Stuttgart Airport & 
EDDV: Hannover Airport \\ \hline
EDDW: Bremen Airport & 
EDSB: Karlsruhe/Baden-Baden Airport \\ \hline
EETN: Lennart Meri Tallinn Airport & 
EFHK: Helsinki-Vantaa Airport \\ \hline
EFOU: Oulu Airport & 
EFRO: Rovaniemi Airport \\ \hline
EFTP: Tampere-Pirkkala Airport & 
EFTU: Turku Airport \\ \hline
EFVA: Vaasa Airport & 
EGBB: Birmingham Airport \\ \hline
EGJJ: Jersey Airport & 
EGNX: East Midlands Airport \\ \hline
EGPD: Aberdeen International Airport & 
EGPF: Glasgow Airport \\ \hline
EHBK: Maastricht Aachen Airport & 
EHRD: Rotterdam The Hague Airport \\ \hline
EINN: Shannon Airport & 
EKAH: Aarhus Airport \\ \hline
EKBI: Billund Airport & 
EKYT: Aalborg Airport \\ \hline
EPGD: Gda\'nsk Lech Wa{\l}{\c e}sa Airport & 
EPKK: Krak\'ow John Paul II International Airport \\ \hline
EPKT: Katowice Wojciech Korfanty Airport & 
EPPO: Pozna\'n-{\L}awica Henryk Wieniawski Airport \\ \hline
EPWR: Wroc{\l}aw Airport & 
ESGG: G\"oteborg Landvetter Airport \\ \hline
ESMS: Malm\"o Airport & 
EVRA: RIX Riga Airport \\ \hline
EYKA: Kaunas Airport & 
EYVI: Vilnius International Airport \\ \hline
LATI: Tirana International Airport N\"en\"e Tereza & 
LBBG: Burgas Airport \\ \hline
LBSF: Sofia Airport & 
LBWN: Varna Airport \\ \hline
LCLK: Larnaca International Airport & 
LCPH: Paphos International Airport \\ \hline
LDDU: Dubrovnik Ruder Bo\v{s}kovi\'c Airport & 
LDPL: Pula Airport \\ \hline
LDSP: Split Saint Jerome Airport & 
LDZA: Zagreb Franjo Tudman Airport \\ \hline
LDZD: Zadar Airport & 
LEAL: Alicante-Elche Miguel Hern\'andez Airport \\ \hline
LEBB: Bilbao Airport & 
LEGE: Girona-Costa Brava Airport  \\ \hline
LEIB: Ibiza Airport & 
LEJR: Jerez Airport \\ \hline
LEST: Santiago-Rosal\'ia de Castro Airport & 
LEXJ: Santander Airport \\ \hline
LEZG: Zaragoza Airport & 
LEZL: Seville Airport \\ \hline
LFBD: Bordeaux-M\'erignac Airport & 
LFBO: Toulouse-Blagnac Airport \\ \hline
LFBZ: Biarritz Pays Basque Airport & 
LFLL: Lyon-Saint-Exup'ery Airport \\ \hline
LFML: Marseille Provence Airport & 
LFMN: Nice C\^ote d'Azur Airport \\ \hline
LFOB: Beauvais-Till\'e Airport & 
LFQQ: Lille Airport \\ \hline
LFRB: Brest Bretagne Airport & 
LFRS: Nantes Atlantique Airport \\ \hline
LFSB: EuroAirport Basel Mulhouse Freiburg & 
LGSA: Chania International Airport Daskalogiannis \\ \hline
LHDC: Debrecen International Airport & 
LICC: Catania-Fontanarossa Airport \\ \hline
LIME: Orio al Serio International Airport & 
LIRP: Pisa International Airport \\ \hline
LJLJ: Ljubljana Jo\v{z}e Pu\v{c}nik Airport & 
LKTB: Brno-Tu\v{r}any Airport \\ \hline
LMML: Malta International Airport & 
LOWG: Graz Airport \\ \hline
LOWI: Innsbruck Airport & 
LOWK: Klagenfurt Airport \\ \hline
LOWL: Linz Airport & 
LOWS: Salzburg Airport \\ \hline
LPMA: Madeira Airport & 
LPPR: Porto Airport \\ \hline
LRCL: Avram Iancu Cluj International Airport & 
LRIA: Ia{\c s}i International Airport \\ \hline
LROP: Henri Coanda International Airport & 
LRTR: Timi{\c s}oara Traian Vuia International Airport \\ \hline
LSZA: Lugano Airport & 
LSZB: Bern-Belp Regional Aerodrome \\ \hline
LTAC: Ankara Esenbo{\u g}a Airport & 
LTAF: Adana {\c S}akirpa{\c s}a Airport \\ \hline
LTAI: Antalya Airport & 
LTBA: Atat\"urk Airport \\ \hline
LTBJ: \.{I}zmir Adnan Menderes Airport & 
LTBS: Dalaman Airport \\ \hline
LTCG: Trabzon Airport & 
LTFE: Milas-Bodrum Airport \\ \hline
LTFJ: Istanbul Sabiha G\"ok{\c c}en International Airport & 
LTFM: Istanbul Airport \\ \hline
LUKK: Chi{\c s}in{\u a}u International Airport & 
LWSK: Skopje International Airport \\ \hline
LYBE: Belgrade Nikola Tesla Airport & 
LZIB: Bratislava Airport \\ \hline

\end{tabular}
}
\end{table*}

\subsection{Air transport delay data}

Time series representing the evolution of delays in the US air transport system have been reconstructed using data obtained from the Reporting Carrier On-Time Performance database of the Bureau of Transportation Statistics, U.S. Department of Transportation, freely accessible at \url{https://www.transtats.bts.gov}. This database contains information about flights operating in US airports, including departure and arrival time (both scheduled and executed), and consequently the associated delays. Data here considered cover years 2015 to 2019 (both included); note that more recent data, while available, have been disregarded, due to the abnormal dynamics experienced during the COVID-19 pandemics.
From this data set, the arrival delay of each flight has been estimated, as the difference between the corresponding actual and scheduled landing times. Afterwards, a time series of hourly average delay has been calculated for each of the top-$30$ airports in terms of number of flights - see Tab. \ref{tab:USDelay}.

\begin{table*}[!tb]
\caption{\label{tab:USDelay} List of the $30$ largest US airports composing the air transport delay data. }
{\small
\begin{tabular}{|p{8.0cm}|p{8.0cm}|}
    \hline

    Hartsfield-Jackson Atlanta International Airport & 
    Denver International Airport \\ \hline
    Dallas Fort Worth International Airport & 
    Los Angeles International Airport \\ \hline
    Chicago O'Hare International Airport & 
    Phoenix Sky Harbor International Airport \\ \hline
    Minneapolis-Saint Paul International Airport & 
    Charlotte Douglas International Airport \\ \hline
    Seattle-Tacoma International Airport & 
    San Francisco International Airport \\ \hline
    John F. Kennedy International Airport & 
    George Bush Intercontinental Airport \\ \hline
    Orlando International Airport & 
    Newark Liberty International Airport \\ \hline
    Harry Reid International Airport & 
    Fort Lauderdale-Hollywood International Airport \\ \hline
    General Edward Lawrence Logan International Airport & 
    Detroit Metropolitan Wayne County Airport \\ \hline
    Miami International Airport & 
    LaGuardia Airport \\ \hline
    Washington Dulles International Airport & 
    Baltimore/Washington International Thurgood Marshall Airport \\ \hline
    Philadelphia International Airport & 
    San Diego International Airport \\ \hline
    Chicago Midway International Airport & 
    Salt Lake City International Airport \\ \hline
    Ronald Reagan Washington National Airport & 
    Tampa International Airport \\ \hline
    Portland International Airport & 
    St. Louis Lambert International Airport \\ \hline
        
\end{tabular}
}
\end{table*}

\subsection{Canadian borders' data}

This data set includes data about the number of daily foreign visitors entering Canada; it is collected by Statistics Canada, i.e. the national statistical office of Canada, and is freely available at \url{https://www150.statcan.gc.ca/t1/tbl1/en/cv.action?pid=2410005601}.
Data have been obtained from January $1^{st}$ 2018 to December $31^{st}$ 2023, and include the daily number of visitors, arriving in any transportation mode, from the $30$ largest incoming countries and regions - a full list is reported in Tab. \ref{tab:CanadaCountries}.

\begin{table*}[!tb]
\caption{\label{tab:CanadaCountries} List of the $30$ regions with the highest number of incoming visitors to Canada. The list include both the name of the region, as defined in the original data set, and the total number of visitors in the considered time window. }
{\small
\begin{tabular}{|p{8.0cm}|p{8.0cm}|}
    \hline

United Kingdom: $2.250.596$ & 
France: $2.214.186$ \\ \hline
Mexico: $1.499.043$ & 
Germany: $940.731$ \\ \hline
Americas countries other than the United States of America - other countries: $382.235$ & 
Netherlands: $367.473$ \\ \hline
Brazil: $352.235$ & 
Switzerland: $337.682$ \\ \hline
Italy: $278.597$ & 
Belgium: $217.295$ \\ \hline
Spain: $213.824$ & 
Ireland: $208.233$ \\ \hline
Europe - other countries: $203.574$ & 
Israel: $136.226$ \\ \hline
Jamaica: $114.370$ & 
Sweden: $110.847$ \\ \hline
Colombia: $94.151$ & 
Denmark: $93.236$ \\ \hline
Austria: $90.445$ & 
Portugal: $87.292$ \\ \hline
Chile: $82.686$ & 
Poland: $79.915$ \\ \hline
Trinidad and Tobago: $71.130$ & 
Norway: $65.322$ \\ \hline
Ukraine: $58.927$ & 
Turkey: $52.242$ \\ \hline
Bahamas: $50.832$ & 
Romania: $50.712$ \\ \hline
Czechia: $49.808$ & 
Bermuda: $46.380$ \\ \hline

\end{tabular}
}
\end{table*}


\newpage
\clearpage

\section{Data preprocessing}
\label{sec:data_pre}

\subsection{EEG data}

For the creation of each network of either a control subject or a PD patient, a random recording from the corresponding group has been selected, for then extracting a segment of length $w$ starting from a random position. The output is then a set of $30$ time series of length $w$, to which we applied an automatic artefact removal procedure based on subtracting the two main components as detected by an Independent Component Analysis. The resulting time series have further been analysed for the presence of silent sequences, suggesting that the data of the channel were not recorded correctly, and for values exceeding $\pm 300$ microvolts; if any such instance was detected, the whole set was discarded, and the process was performed again.

\begin{figure}[b]
\centering
\includegraphics[width=.65\linewidth]{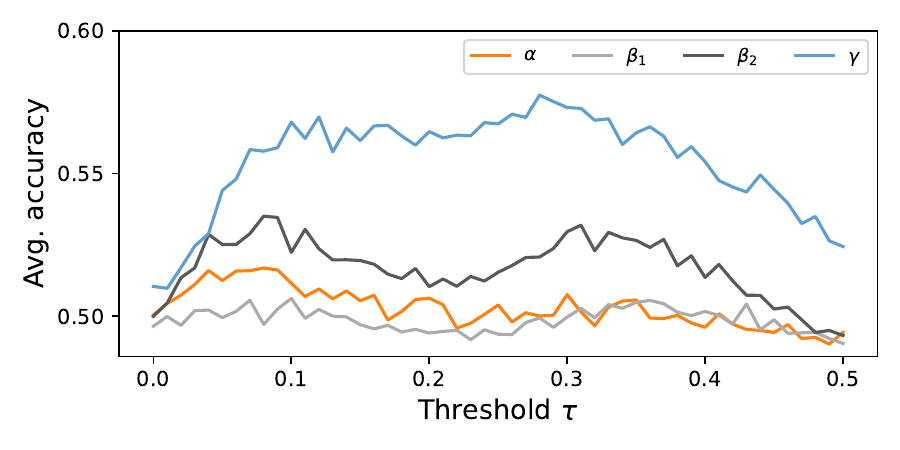}
\caption{ Evolution of the classification score between functional networks of control subjects and patients, when the corresponding time series are filtered according to four standard frequency bands (see legend for colour codes). Results are reported as the average accuracy over $500$ independent realisations, as a function of the threshold $\tau$ - see Sec. \ref{sec:net_rec} for details. }
\label{fig:EEGFreq}
\end{figure}

It is worth noting that all results here presented correspond to the analysis of the broadband signal, as yielded by the EEG machine. For the sake of completeness, we further tested the effect of reconstructing functional networks (see Sec. \ref{sec:net_rec}) on signals filtered according to four standard frequency bands, namely $\alpha$ ($8$ - $13$Hz), $\beta_1$ ($13$ - $20$Hz), $\beta_2$ ($20$ - $30$Hz), and $\gamma$ ($30$ - $50$Hz). In all cases, the processing has been performed with a fourth order Butterworth filter. The resulting networks present much smaller differences between control subjects and patients, as represented by the corresponding classification score in Fig. \ref{fig:EEGFreq}; the frequency band approach has then been discarded.

\subsection{Stock markets' data}

To create the stock market's networks, we selected the adjusted close price daily time series for the companies composing the chosen market. These raw time series start on the day of the opening of the youngest company on the market. In the case of Euro Stoxx 50, holidays in different countries where companies remained closed were removed from the time series. Then, for every time series, a logarithmic normalisation is applied to obtain a representation of the relative increase or decrease in the adjusted closed price with respect to the previous day. As customary, this is done through the transformation:

\begin{equation}
    \label{eq:log_norm}
    \hat{x}(t) = \log_2 \frac{ x(t) }{ x(t-1) }.
\end{equation}

The output is therefore 30 normalised time series for each market described in the previous section (Sec. \ref{sec:data_sources}).

\subsection{Forex' data}

Forex' time series were characterised by some missing data, which especially affected the beginning and the end of each year and major international holidays. These were dealt with by firstly associating a UNIX time stamp to each available data point, i.e. a number representing the elapsed time in seconds since a reference moment; secondly, missing points were recovered using a piece-wise linear interpolation. As a final step, time series have been transformed according to the logarithm of the ratio between consecutive time points, see Eq. \ref{eq:log_norm}. No additional processing has been performed.

\subsection{Weather (wind) data}

The wind speed data obtained for each airport have firstly been converted to knots, in those few cases in which they were reported in other units (e.g. meters per second). 
Secondly, due to both internal technical limitations (e.g. connectivity losses) and to the incompleteness of the original data, a small fraction of time points were missing. Note that airports were initially filtered, and those for which more than $10\%$ of the reports were missing, or for which more than five times data were missing in consecutive gaps of $6$ or more hours, were already excluded. Hence, the few remaining instances of missing data have been solved by resorting to a linear interpolation.

\subsection{Air transport delay data}

Time series representing the dynamics of air transport are highly non-stationary, i.e. their evolution is a function of time; to illustrate, delays mostly appear whenever there is an increase in the number of operations, and are thus larger during the mid-day peak hours, for then substantially dropping in late hours of the day. Note that the presence of such regular trends may lead to an overestimation of the functional connectivity between airports. To solve this problem, each time series $x(t)$ is transformed to the corresponding Z-Score, defined as:

\begin{equation}
\label{eq:zscore}
    z(t) = \frac{ x(t) - \overline{x(t+24k)} }{ \sigma_{x(t+24k))} },
\end{equation}

with $k = ( \ldots, -2, -1, 1, 2, \ldots)$, $\overline{ \cdot }$ being the average operator, and $\sigma_{ \cdot }$ the standard deviation. In other words, each value $z(t)$ represents the deviation of the observed metric from what expected at the same time in neighbouring days. Regularities appearing at specific time windows are therefore removed, and the resulting time series are highly stationary.

\subsection{Canadian borders' data}

The time series obtained from Statistics Canada have been transformed according to the logarithm of the ratio between consecutive days, i.e. following the same procedure for pre-processing financial data - see Eq. \ref{eq:log_norm}. No additional processing has been performed.


\newpage
\clearpage

\section{Network reconstruction}
\label{sec:net_rec}

For each complex system and each data set previously described in Sec. \ref{sec:data_pre}, the available information includes a set of $30$ time series, which are the starting point for the reconstruction of the corresponding functional networks. Given a window length $w$ and a functional metric (i.e. linear correlation, rank correlation, Granger Causality or Transfer Entropy), the same random segment of length $w$ is selected from the $30$ time series, and the connectivity between all pairs of them is computed. For the sake of completeness, a short description of these functional metrics is here provided:

\begin{itemize}

    \item Linear Correlation (LC). Pearson's linear correlation between the two considered time series.

    \item Rank Correlation (RC). Spearman's rank correlation between the two analysed time series.
    
    \item Granger Causality (GC). A classical example of a {\it predictive causality} test \citep{diebold1998elements}. GC is based on assessing whether the inclusion of information about the dynamics of the causing element helps predict the future dynamics of the caused one \citep{granger1969investigating}. The test is based on comparing two forecast models, usually constructed around autoregressive-moving-average (ARMA) models, respectively including or not information about the past of the causing element; for finally performing an F-test on the residuals, and thus obtaining a $p$-value assessing whether the causing element contributes with relevant information - and hence, whether a causality relationship is present.

    \item Transfer Entropy (TE). TE can be seen as a generalisation of GC based on information-theoretic concepts; and aims at quantifying the reduction in uncertainty of a future value of the caused dynamics due to the history of the cause. In other words, TE evaluates the extra information that the historical state of the cause provides about the next observation in the caused element. The concept of Transfer Entropy was pioneered independently by Schreiber \cite{schreiber2000measuring} and Palu{\v{s}} \cite{paluvs2001synchronization}, and is mathematically formalised as:

    \begin{equation}
	TE_{X\rightarrow Y(k, l)} = 
    \sum_{x \in X, y \in Y} p(y_{n+1}, y^{(l)}_n, x^{(k)}_n) \log \left( \frac{p(y_{n+1}|y^{(l)}_n, x^{(k)}_n)}{p(y_{n+1}|y^{(l)}_n)} \right).
	\end{equation}
 
	$l$ and $k$ denote the embedding vectors, i.e. the past values of the $Y$ and $X$ processes, such that for instance $x^{(k)}_n$ corresponds to $(x_n, \ldots, x_{n-k+1})$. By capturing the direct exchange of information between two variables and being sensitive to non-linear attributes, TE is one of the most powerful measures for functional network reconstruction \cite{vicente2011transfer}.
    
\end{itemize}

In the case of the LC and the RC, we account for the time delay in the information transmission between two time series by finding the maximum value obtained by shifting one of the time series relative to the other by up to $5$ time points. Note that this process is inherently included in the GC and the TE, and is not needed.

The result of this analysis is a connectivity value for each pair of time series, which can be interpreted as the weighted adjacency matrix of a network, where nodes represent time series and edge weights the connectivity (either correlation or causality) between them. The result is thus a complete weighted network. 

As a final step, these complete networks are pruned, by applying a threshold $\tau$ to the edges. This is used to create two different networks:

\begin{itemize}
    \item Unweighted networks: all links with a weight lower than $\tau$ are deleted, and the weight of the surviving links is further deleted - thus creating a sparse unweighted network.
    \item Weighted networks: links whose weight is lower than $\tau$ are deleted, but the weight of the remainder links is preserved; the result is therefore a sparse network with weighted links.
\end{itemize}

\newpage
\clearpage

\section{Network optimisation}

\subsection{General procedure}

As previously described, the reconstruction of a functional network relies on two parameters: the segment length $l$ and the threshold $\tau$ - note that a third parameter, representing a temporal downsampling of the data, will be further studied in Sec. \ref{sec:downsampling}. We here discuss how the best combination of parameters has been obtained in each case; note that the same procedure has independently been applied to both weighted and unweighted networks.

\subsubsection*{EEG brain networks}

The procedure starts by fixing $l = 1,000$, both for networks corresponding to control subjects and patients suffering from Parkinson's Disease; the threshold $\tau$ is then varied between $0.0$ and $0.5$, in steps of $0.01$. The best $\tau$, i.e. the one for which the classification score is maximal, is retained; and afterwards $l$ is varied between $100$ and $2,000$, in steps of $100$, again retaining the value yielding a maximal classification score.

The same procedure is independently applied to optimise the network reconstruction for the four functional metrics previously described; and for data filtered in specific frequency bands, as shown in Fig. \ref{fig:EEGFreq}.

\subsubsection*{S\&P 500, Forex, winds, air delays, Canadian borders}

The optimisation procedure is also a greedy one, similar to the previous case: the threshold $\tau$ is initially fixed to $0.2$, and the segment length $l$ is varied (in this case, the range has been independently estimated for each data set, see the corresponding graphs); once the best $l$ is obtained, $\tau$ is varied between $0.0$ and $0.5$. The main difference is that, in this case, we consider that the parameter is optimal when it minimises the classification score between the considered networks and the EEG ones, as the objective is to find comparable representations.

\subsubsection*{Additional stock markets}

In order to evaluate the generalisability of the parameters previously obtained, the following additional stock markets have been considered: DJI, Euro Stoxx 50, Aggressive Small Caps, DAX, FTSE, HSI, KLSE, and STI - see Sec. \ref{sec:data_sources} for details. In the case of these data sets, the networks have been reconstructed using the same parameters as the ones obtained for the S\&P 500 one; i.e., no further optimisation has been performed.

\subsubsection*{Extended data sets}

In the case of Forex, NASDAQ 500k, and winds, more time series were available, beyond the $30$ initially considered. In order to assess the generalisability of the procedure, $2,000$ random sets of time series have been extracted from each data set, and the corresponding functional networks have been reconstructed using the same parameters as the initial case. Results are reported in Fig. 3 of the main manuscript.

\newpage
\clearpage

\subsection{Optimisation of winds' networks}

The optimisation and reconstruction of winds' networks have followed the same procedures as all other complex systems; the results have nevertheless been omitted from Fig. 1 in the main manuscript, and are here reported instead for the sake of clarity. Specifically, Fig. \ref{fig:Winds_opt} below depicts the evolution of the classification score as a function of the segment length $l$ and of the applied thresholds $\tau$. While results for weighted networks (see bottom panels) are comparable to what obtained for other complex systems, unweighted networks (top panels) present a minimum for very long time scales. In order to reduce the impact of the noise, we calculated a second-order polynomial fit of the curve between $l = 600$ and $1300$ (see black solid line in the left panel); and the optimal time series length has been defined as the minimum of that fit. The impact of the use of such long segments, and specifically the relationship of this with the system characteristic time scale, will be further discussed in Sec. \ref{sec:downsampling} below.

\vspace{2cm}

\begin{figure}[h]
\centering
\includegraphics[width=.8\linewidth]{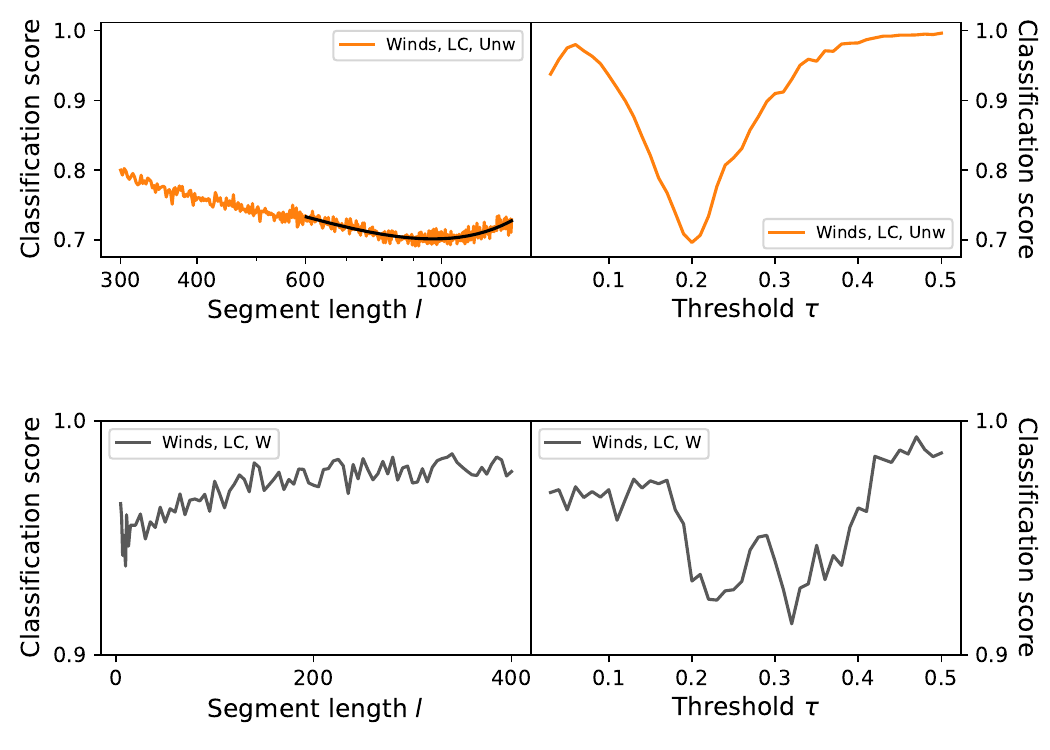}
\caption{ Classification score between between functional brain networks of control subjects and those obtained from wind data, both unweighted (top panels) and weighted (bottom panels), as a function of the segment length $l$ (left panels) and the threshold $\tau$ (right panel). The black solid line in the top left panel reports the best fit of a second-order polynomial function, between $l = 600$ and $1300$. }
\label{fig:Winds_opt}
\end{figure}

\newpage
\clearpage

\subsection{Optimality of the optimisation process}

\begin{figure}[h]
\centering
\includegraphics[width=.35\linewidth]{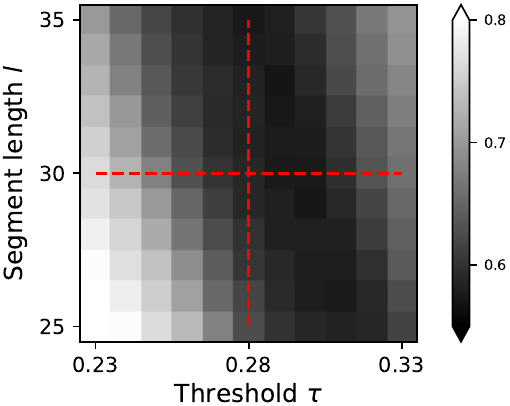}
\caption{ Classification score of Forex networks as a function of the segment length $l$ and the threshold $\tau$. Red lines indicate the solution found by the greedy optimisation, i.e. $l = 28$ and $\tau = 0.3$. }
\label{fig:Tuning}
\end{figure}

The attentive reader will have noted that the optimisation process here used is a greedy one, in which the impact of changing the two parameters is evaluated sequentially. In order to check that this is not biasing the results, Fig. \ref{fig:Tuning} reports an analysis of the classification score, for Forex functional networks, around the optimal parameter set previously identified (i.e. $l = 28$ and $\tau = 0.3$). It can be appreciated that the solution obtained (with a classification score of $0.588$) is very close to the global optimum (minimum score of $0.571$); hence the use of a greedy optimisation does not significantly impact results, but allows to substantially reduce the computational cost of the analysis.

\subsection{Synthesis of network reconstruction parameters}

For the sake of clarity, we report in Tab. \ref{tab:SynthParams} a synthesis of the optimal parameters used in the reconstruction of each functional network set, including values of $l$, $\tau$, and the downsampling $\gamma$ (further analysed in Sec. \ref{sec:downsampling}). The fifth column further reports the real time duration of the data segments used for each data set.

\begin{table*}[!b]
\caption{\label{tab:SynthParams} Synthesis of the best parameters used in the reconstruction of each set of functional networks. The column real time indicates the time represented by $l$ in the real evolution of the system. Rows in italic correspond to systems for which the parameters of another one have been used. }
{\small
\begin{tabular}{|p{3.0cm}|p{2.2cm}|p{2.2cm}|p{2.5cm}|p{2.2cm}|p{3.0cm}|}
    \hline

{\bf Data set} & {\bf Filter} & {\bf Length $l$} & {\bf Threshold $\tau$} & {\bf Real time} & {\bf Downsampling $\gamma$} \\ \hline

Brain & Unweighted & 1400 & 0.2 & 2.8s & - \\
 & Weighted & 1400 & 0.29 & 2.8s & - \\ \hline
Forex & Unweighted & 28 & 0.3 & 28 min & 95\\
 & Weighted & 22 & 0.18 & 22 min & - \\ \hline
S\&P500 & Unweighted & 33 & 0.21 & 33 days & - \\
 & Weighted & 31 & 0.2 & 31 days & - \\ \hline
{\it DJI} & {\it Unweighted} & {\it 33} & {\it 0.2} & {\it 33 days} & - \\ \hline
{\it EuroStoxx 500} & {\it Unweighted} & {\it 33} & {\it 0.2} & {\it 33 days} & - \\ \hline
{\it ASC} & {\it Unweighted} & {\it 33} & {\it 0.2} & {\it 33 days} & - \\ \hline
{\it DAX} & {\it Unweighted} & {\it 33} & {\it 0.2} & {\it 33 days} & - \\ \hline
{\it FTSE} & {\it Unweighted} & {\it 33} & {\it 0.2} & {\it 33 days} & - \\ \hline
{\it HSI} & {\it Unweighted} & {\it 33} & {\it 0.2} & {\it 33 days} & - \\ \hline
{\it KLSE} & {\it Unweighted} & {\it 33} & {\it 0.2} & {\it 33 days} & - \\ \hline
{\it STI} & {\it Unweighted} & {\it 33} & {\it 0.2} & {\it 33 days} & - \\ \hline
Winds & Unweighted & 968 & 0.2 & 20 days & 1 \\
 & Weighted & 10 & 0.32 & 5 hours & - \\ \hline
Air delays & Unweighted & 21 & 0.25 & 21 hours & - \\
 & Weighted & 10 & 0.19 & 10 hours & - \\ \hline
Canadian & Unweighted & 32 & 0.26 & 32 days & - \\
 & Weighted & 7 & 0.31 & 7 days & - \\ \hline
 
\end{tabular}
}
\end{table*}

\newpage
\clearpage

\section{Network analysis}

\subsection{Financial networks and market volume}

\begin{figure}[!h]
\centering
\includegraphics[width=.8\linewidth]{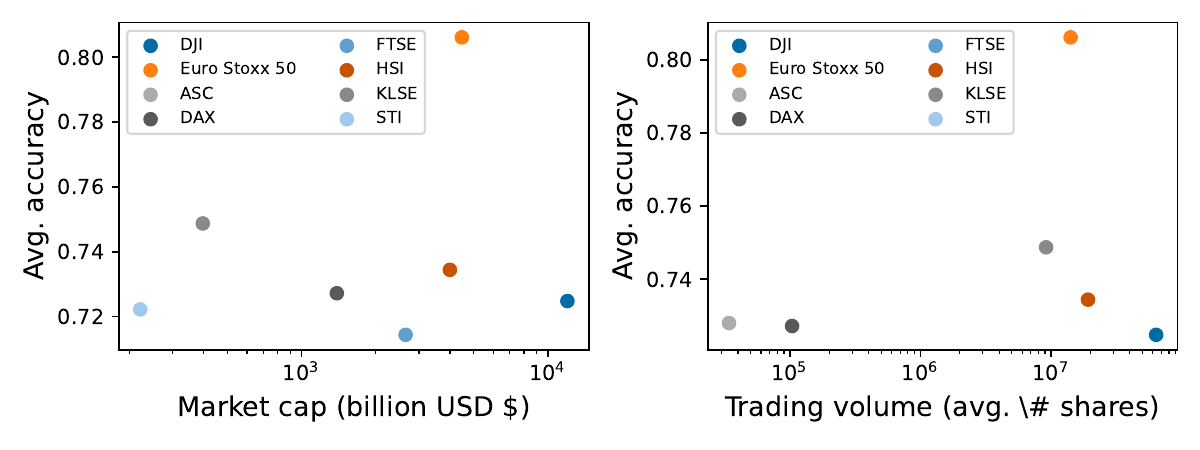}
\caption{ Average classification score for eight considered stock markets, as a function of their market capitalisation (left panel) and the average daily trading volume (right panel). }
\label{fig:Volume}
\end{figure}

As a potential explanation of the differences in identifiability we observe across different stock markets, we here explore the relationship between this and their trading volume. Note that, according to the Efficient Market Hypothesis \cite{malkiel1989efficient, sewell2012efficient, naseer2015efficient}, larger (and more mature) markets should be more efficient, in the sense that information is more evenly distributed among participants, and this should result in clearer computations. Fig. \ref{fig:Volume} reports scatter plots of the average classification score as a function of the market capitalisation (left panel) and of the average trading volume (in number of stocks, right panel). No clear trend can be identified; size can thus be discarded as an explaining factor.

\vspace{2cm}

\newpage
\clearpage

\subsection{Winds' networks and spatial location}

\begin{figure}[h]
\centering
\includegraphics[width=.85\linewidth]{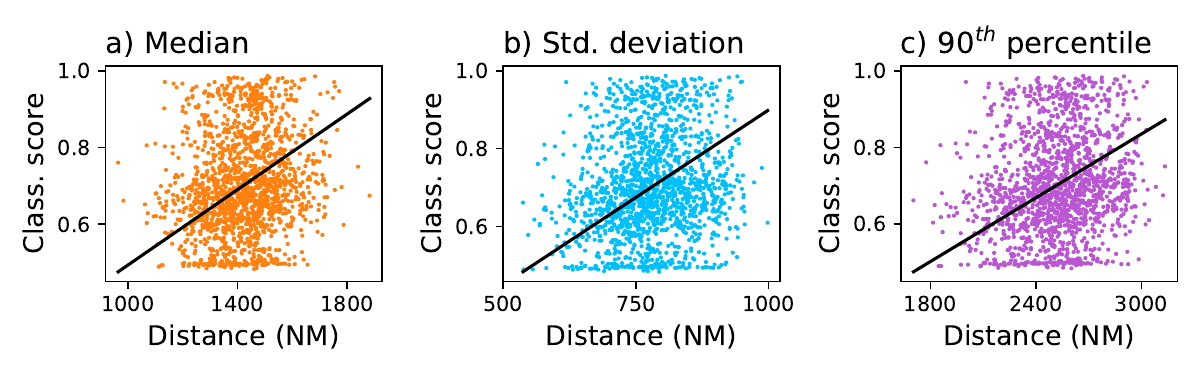}
\caption{ Scatter plots of the classification score obtained for winds' networks, as a function of the median (left), standard deviation (centre), and $90^{th}$ percentile (right) of the distance between the airports in there included. The black lines represent the best linear fit. }
\label{fig:WeatherDist}
\end{figure}

Following the previous case, we here analyse the relationship between the identifiability of winds' networks, and the geographical dispersion of the nodes (i.e. airports) composing them. Using the full list of available airports (see Tab. \ref{tab:EUAirports2}), $2,000$ subsets of $30$ random airports have been selected, for then evaluating their identifiability. The obtained classification score has then been compared to several properties of the distribution of distances between pairs of airports, as reported in Fig. \ref{fig:WeatherDist} and Tab. \ref{tab:MixedDist}.

A linear fit between the score and each property always yields statistically significant results (see Tab. \ref{tab:MixedDist}), although a graphical representation of such fits hardly supports the presence of a relationship (see Fig. \ref{fig:WeatherDist}). These fits suggest that networks are less identifiable (i.e., more similar to EEG ones) when airports are close together, and are further homogeneously distributed.

\vspace{2cm}

\vspace{2cm}

\begin{table*}[!h]
\caption{\label{tab:MixedDist} Details of the linear fits between the classification score obtained for winds' networks, and seven metrics extracted from the distribution of distances between pairs of airports in them included. }
{\small
\begin{tabular}{|p{4.0cm}|p{2.0cm}|p{2.0cm}|p{2.0cm}|p{2.0cm}|}
    \hline
{\bf Statistics} & {\bf Slope} & {\bf Intercept} & {\bf $R^2$} & {\bf $p$-value} \\ \hline
Mean & 0.000168 & 0.466 & 0.169 & $< 8 \cdot 10^{-12}$ \\
Median & 0.000136 & 0.525 & 0.149 & $< 10^{-9}$ \\
Standard deviation & 0.000266 & 0.497 & 0.161 & $< 7 \cdot 10^{-11}$ \\
Minimum & 0.000310 & 0.685 & 0.064 & $< 10^{-2}$ \\
Maximum & 0.000027 & 0.601 & 0.100 & $< 5 \cdot 10^{-5}$ \\
$10 ^{th}$ percentile & 0.000242 & 0.587 & 0.111 & $< 8 \cdot 10^{-6}$ \\
$90 ^{th}$ percentile & 0.000099 & 0.455 & 0.180 & $< 5 \cdot 10^{-13}$ \\ \hline

\end{tabular}
}
\end{table*}

\newpage
\clearpage

\subsection{Topological metrics}

In order to understand whether the similarity observed between functional networks representing the brain activity of control subjects and the Forex ones is biased by the use of a DL model, we here compare the topological properties of both sets of networks. These are metrics that describe specific aspects of the structure of a network, and therefore allow to synthesise it in single values. For each network we calculated the following six standard metrics:

\begin{itemize}
    \item {\it Assortativity}: metric measuring the tendency of the nodes of a network to be connected to nodes with degree (number of connections) very similar to it \cite{noldus2015assortativity}. It is calculated as the Pearson's correlation coefficient between the degree of nodes at both end of each link.
    \item {\it Transitivity}: tendency of the network to form triangles, i.e. triplets of nodes connected between them, normalised by the total number of possible triplets \cite{serrano2006clustering}.
    \item {\it Efficiency}: calculated as the mean inverse distance between pairs of nodes of the network, it indicates how easily information can move between nodes \cite{latora2001efficient}.
    \item {\it Modularity}: tendency of the network of organising in communities, i.e. groups of nodes strongly connected between them but loosely connected with the remainder ones \cite{newman2006modularity, fortunato2010community}. The community structure is here estimated using the Louvain algorithm \cite{blondel2008fast}.
    \item {\it S Metric}: metric describing the presence of a ``hub-like core'', where hubs (i.e. nodes with a large number of connections) are also connected between themselves to form a core \cite{li2005towards}. It is defined as:
    \begin{equation}
        s = \sum_{i,j} d_i d_j,
    \end{equation}
    with $d_i$ being the degree (or number of links) of node $i$, and $i$ and $j$ representing the end nodes of all links in the network.
    \item {\it Network Information Content} (NIC): measurement of the presence of regularities in the adjacency matrix, and hence of the amount of information encoded within the structure of the network \cite{zanin2014information}. It is calculated by merging pairs of nodes that imply the lowest loss of information, until shrinking the network into a single node.
\end{itemize}

All these metrics, with the exception of the modularity, depend on the number of nodes and of links in the network, and not only on the topological structure they aim at describing. To illustrate, the larger the number of links in the network, the easier is to obtain triangles by chance. In order to normalise these metrics, and thus allow comparisons between networks of different sizes, we resort to a null model composed of random equivalent (same number of nodes and links) networks. The original metric is then expressed through the corresponding Z-Score, which then represents how an observed metric deviates from randomness.

\begin{figure}[t]
\centering
\includegraphics[width=.55\linewidth]{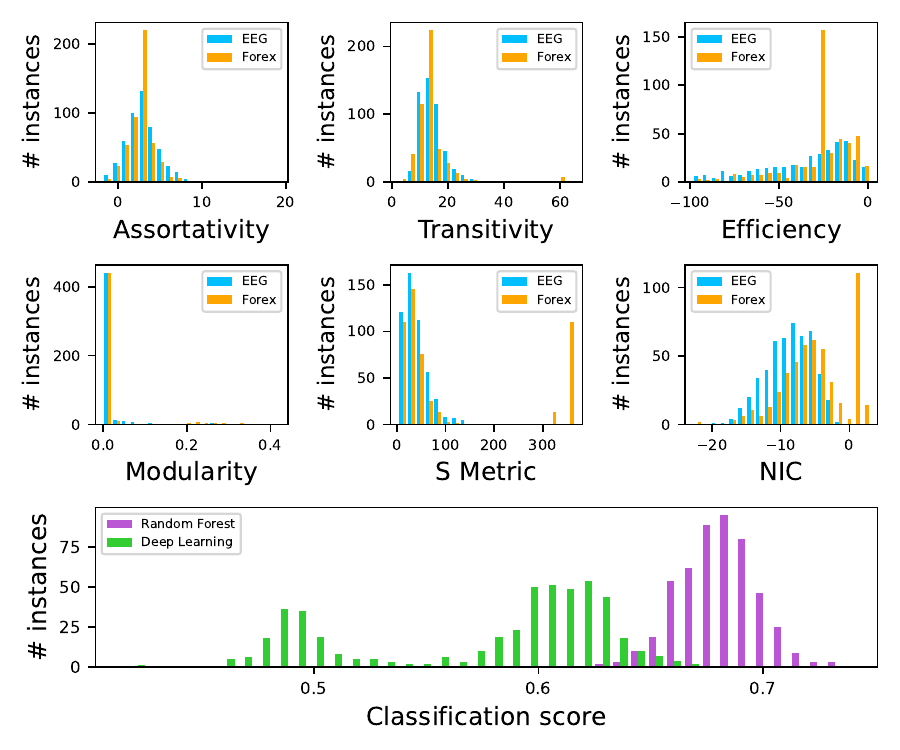}
\caption{ Comparing functional networks of control subjects and Forex. The top six panels report histograms of the values of topological metrics of these networks, for EEG (blue bars) and Forex (orange bars) networks; see text for metric definitions. The bottom panel reports the distributions of the classification score obtained by Random Forest (magenta bars) and Deep Learning (green bars) models, the former being trained on the aforementioned topological metrics. }
\label{fig:EEG_Forex}
\end{figure}

Each of the top six panels of Fig. \ref{fig:EEG_Forex} reports two histograms, comparing the probability distribution of the metric values for the unweighted functional networks of healthy control subjects (blue) and Forex (orange). No pair of distributions are exactly the same - a two-sample Kolmogorov-Smirnov test only accepts equality for the modularity for a significance level $\alpha = 0.01$. Still, most metrics are very similar in both cases, and the differences are not enough to classify individual networks.

In order to exclude that those differences could actually be used in a classification task, we have further trained a Random Forest (RF) model with such topological features and evaluated its performance. The RF is a classical Machine Learning model widely used in classification tasks, which provides both good accuracy and low tendency to overfitting - see Ref. \cite{breiman2001random} for an in-depth description. Results here correspond to a set of $500$ tasks; in each one of them, a random half of the data was used for training, and the remainder for testing. The final score has then been assessed through the corresponding accuracy, as in the case of the DL classification. The bottom panel of Fig. \ref{fig:EEG_Forex} indicates that the RF model trained over topological features is in general better than the DL one, and is further more stable - note how the latter at times does not converge to a solution, hence the bimodal distribution. Still, the median score obtained by the RF model ($68.6\%$) is not much higher than the DL's one, and still supports the main arguments of this work.

\begin{figure}[b]
\centering
\includegraphics[width=.55\linewidth]{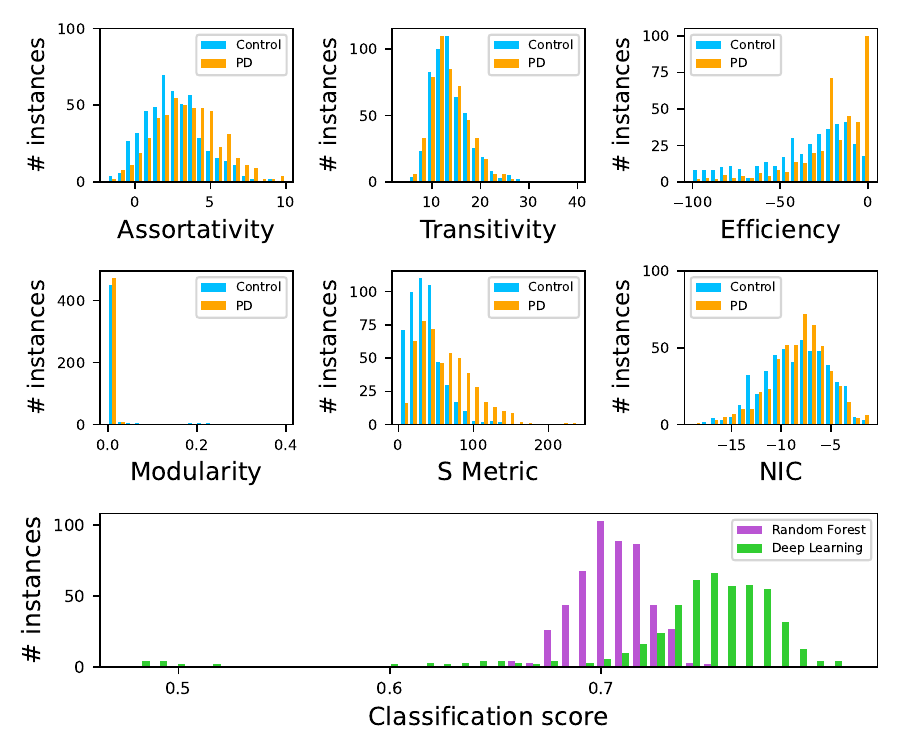}
\caption{ Comparing functional networks of control subjects and Parkinson's patients. The top six panels report histograms of the values of topological metrics of these networks, for control subjects' (blue bars) and PD patients' (orange bars) networks; see text for metric definitions. The bottom panel reports the distributions of the classification score obtained by Random Forest (magenta bars) and Deep Learning (green bars) models, the former being trained on the aforementioned topological metrics. }
\label{fig:EEG_EEG}
\end{figure}

We performed the same analyses on the functional networks of control subjects and PD patients, in order to recover the main classification task presented in the manuscript. As can be seen in Fig. \ref{fig:EEG_EEG}, differences in topological metrics are minimal, and this results in a limited capacity of the RF model to discriminate between the two groups. 

In short, results obtained by the DL model are qualitatively similar to those that could be obtained by standard Machine Learning models trained on the topological properties of the networks. The reasons behind the difference between them, and specifically why the RF model outperforms DL in Fig. \ref{fig:EEG_Forex}, are open questions at the moment.

\newpage
\clearpage

\subsection{Motifs}

Motifs, i.e. patterns of connectivity involved a limited number of nodes that appear with a frequency higher than what expected at random, have frequency been associated with computation of information in networks. We therefore tried to understand whether the appearance of motifs is somehow related with the similarity with EEG networks.

\begin{figure}[h]
\centering
\includegraphics[width=.85\linewidth]{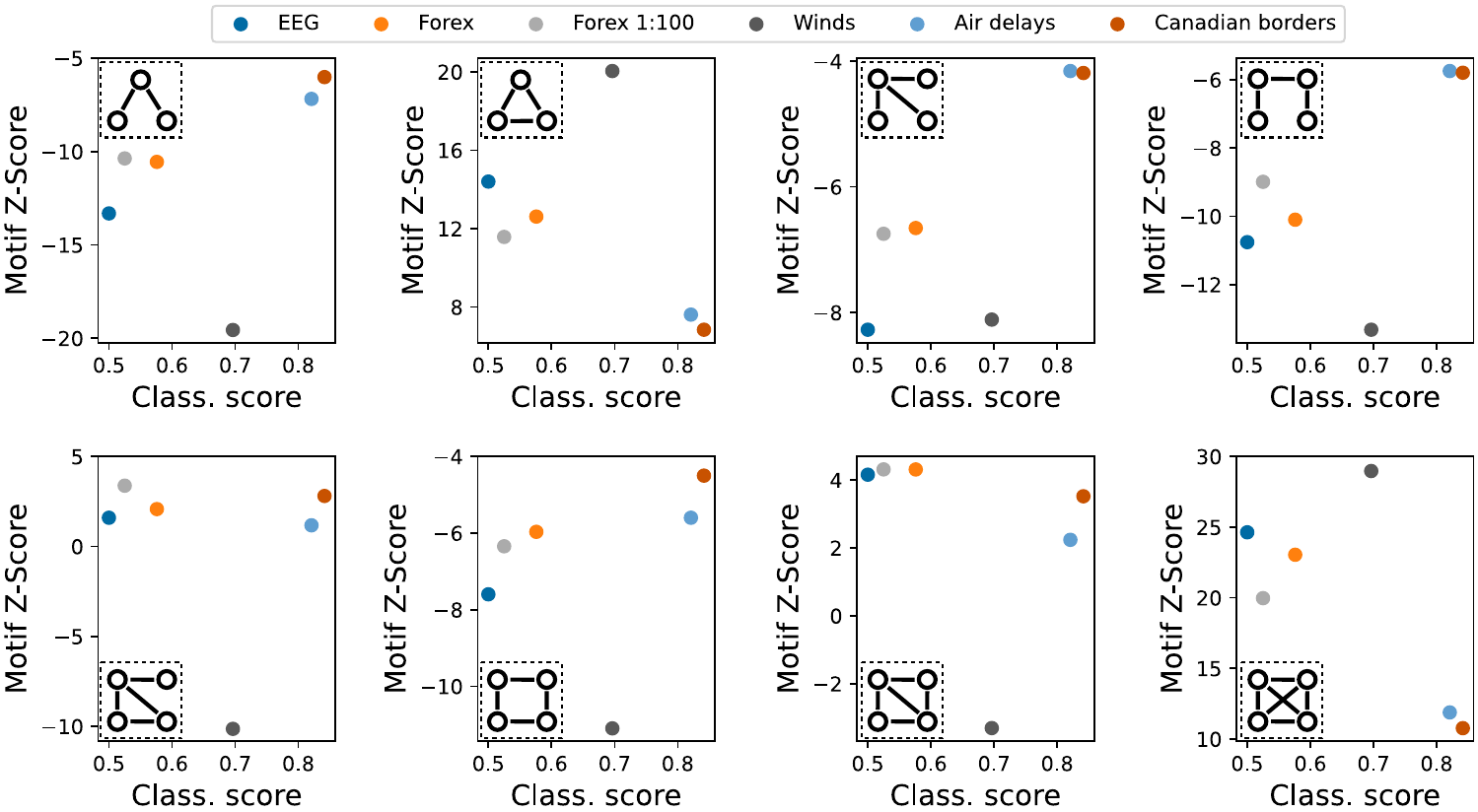}
\caption{ Scatter plot of the Z-Score of $3$- and $4$-nodes motifs as a function of the classification score, as observed in sets of $1000$ networks representing the complex systems reported in the legend. Networks have been reconstructed according to the optimal classification score. }
\label{fig:Motifs}
\end{figure}

Towards this aim, Fig. \ref{fig:Motifs} reports the Z-Score of $3$- and $4$-nodes undirected motifs as a function of the observed classification score. The Z-Score has been calculated as the number of standard deviations the observed number of instances deviates from what expected in ensembles of $10^3$ random equivalent (i.e. same number of nodes and links) networks. We further report results for the EEG functional networks of control subjects; the top-$30$ Forex networks; the same Forex networks, when time series have been downsampled $1:100$; and winds, air delays, and Canadian borders networks - see top legend in Fig. \ref{fig:Motifs}.

The motifs showing the higher appearance frequency are the fully connected triplets and quadruplets of nodes; on the other hand, large negative Z-Scores are associated with incomplete triplets (i.e. lines) and quadruplets (i.e. rings). These results are in line with that presented in the previous section, and especially the high values observed for the S metric (see Figs. \ref{fig:EEG_Forex} and \ref{fig:EEG_EEG}); and suggest the presence of core groups of nodes strongly connected between themselves, creating a core-periphery configuration. Still, the presence of such structure is not enough to justify the similarity between networks, as illustrated by the similar motif values obtained in the case of air delays and Canadian border networks.

\vspace{3cm}

\newpage
\clearpage

\section{Sensitivity analysis}

With the aim to assess whether low classification scores are really significant, and not just a lack of sensitivity of the DL model, we analyse how these scores are affected by controlled changes in the network topology. We specifically consider the set of unweighted functional networks obtained for healthy control subjects, and a second set of these with some properties being modified; for finally performing a classification between them.

\begin{figure}[h]
\centering
\includegraphics[width=.95\linewidth]{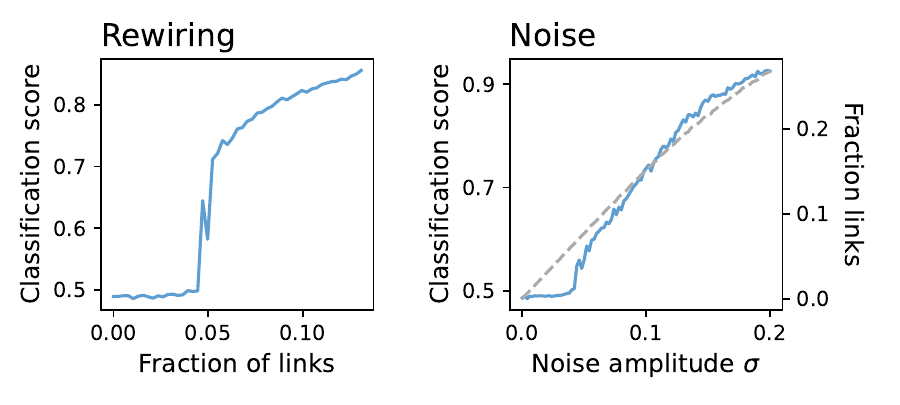}
\caption{ Evolution of the classification score when comparing unweighted networks with slightly altered topology, as a function of the fraction of links randomly rewired (top panel), and of the amplitude of a uniformly distributed observational noise (bottom panel). In the latter case, the dashed grey line (right Y axis) reports the evolution of the fraction of altered links.  }
\label{fig:Sensitivity}
\end{figure}

The first scenario consists of a random rewiring, in which two pairs of nodes $n_{1 \ldots 4}$ are randomly selected, such that a link between $n_1$ and $n_2$ exists, but $n_3$ and $n_4$ are disconnected. We then proceed to move the link from $n_1 - n_2$ to $n_3 - n_4$, thus achieving a slightly modified topology. The left panel of Fig. \ref{fig:Sensitivity} reports the evolution of the median classification score as a function of the fraction of links modified.

As a second scenario, we add a normally distributed noise to the original time series before reconstructing the functional networks - effectively driving the topology towards a random structure. The right panel of Fig. \ref{fig:Sensitivity} reports the evolution of the classification score as a function of the standard deviation $\sigma$ of such noise, normalised against the standard deviation of the corresponding EEG time series. As a reference, the dashed grey line (right Y scale) depicts the corresponding fraction of links that have changed, with respect to the networks reconstructed without noise.

A transition can be observed in both cases, with the score substantially increasing when more than $5\%$ of the links are perturbed. It can therefore be concluded that the DL model is quite sensitive to changes in the structure, as very similar networks are needed to obtain low classification scores.

\vspace{3cm}

\newpage
\clearpage

\section{Temporal downsampling}
\label{sec:downsampling}

Any dynamical system is expected to have one or multiple characteristic time scales, on which the evolution mainly takes place. While these have been extensively studied in the human brain \cite{koenig2005brain, kiebel2008hierarchy, liegeois2019resting, golesorkhi2021brain}, fewer is known for other complex systems, and especially the ones here considered. Consequently, we here consider a downsampling of the original time series, for the Forex and Winds systems - i.e. the two with the higher timer resolution. In both cases, one every $\gamma$ points in the original data have been preserved, while discarding all others. Besides this, the reconstruction has been performed using the best parameters (i.e. time segment lengths and thresholds) previously obtained in each case.

\begin{figure}[h]
\centering
\includegraphics[width=.75\linewidth]{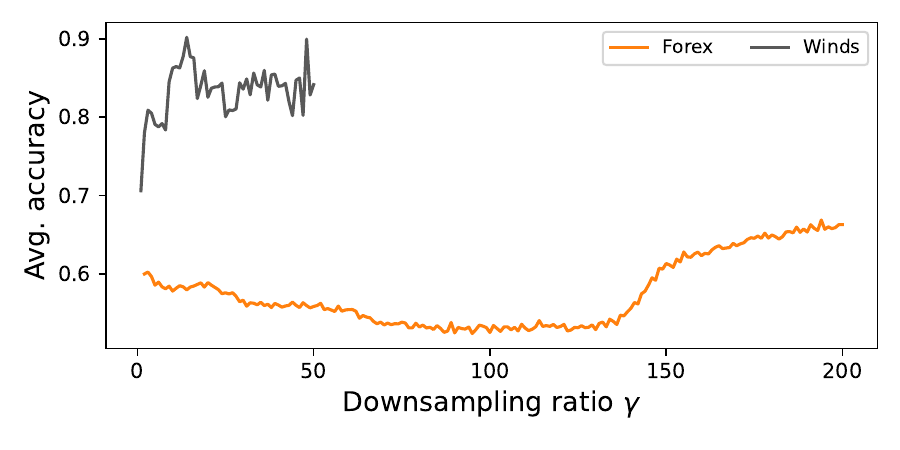}
\caption{ Evolution of the classification score as a function of the downsampling applied to the original time series. Orange and grey lines respectively correspond to Forex and Winds functional networks. The left-most point, i.e. $\gamma = 1$, represent the value obtained with the original data.  }
\label{fig:DownSampling}
\end{figure}

The evolution of the classification score as a function of the downsampling factor $\gamma$ is presented in Fig. \ref{fig:DownSampling}. The classification of the winds' networks improves with all downsampling, i.e. they become more identifiable with respect to EEG ones. On the contrary, in the case of Forex' networks, the classification score drops to a minimum of $0.524$ for $\gamma \approx 95$, corresponding to approximately one data point every $1.5$ hours.

\vspace{2cm}

\newpage
\clearpage

\section{Network data set}

With the aim of fostering future analyses on this topic, we provide the community with a public data set of functional networks. The file {\it FunctionalNetworks.zip} includes $10^4$ unweighted networks, half of them representing the brain activity of control subjects, and the remainder the best optimisation obtained for Forex networks (i.e. $l = 30$, $\tau = 0.31$, and $\gamma = 95$). Files are text encoded, with each row representing a link, and with two tab-separated columns respectively reporting the source and destination nodes. Nodes are represented by integer numbers in range $[0, 29]$. An example of the structure is reported in Tab. \ref{tab:ExampleFile} below.

\begin{table*}[!h]
\caption{\label{tab:ExampleFile} Example of the content of the first network file, i.e. {\it Links\_EEG\_0000.txt}. Note that the first column (in italic) is only included here for reference, and is not actually contained in the file. }
{\small
\begin{tabular}{|p{2.5cm}|p{3.0cm}|p{3.0cm}|}
    \hline
{\it Row number} & {\bf Source node} & {\bf Target node} \\ \hline
{\it 1} & 3 & 4 \\
{\it 2} & 3 & 7 \\
{\it 3} & 3 & 8 \\
{\it 4} & 3 & 9 \\
{\it 5} & 3 & 13 \\
{\it 6} & 3 & 14 \\
{\it 7} & 3 & 18 \\
{\it 8} & 3 & 21 \\
{\it 9} & 4 & 5 \\
{\it 10} & 4 & 7 \\ 
$\ldots$ & $\ldots$ & $\ldots$ \\ \hline 
\end{tabular}
}
\end{table*}

\vspace{2cm}

\begin{figure}[h]
\centering
\includegraphics[width=.95\linewidth]{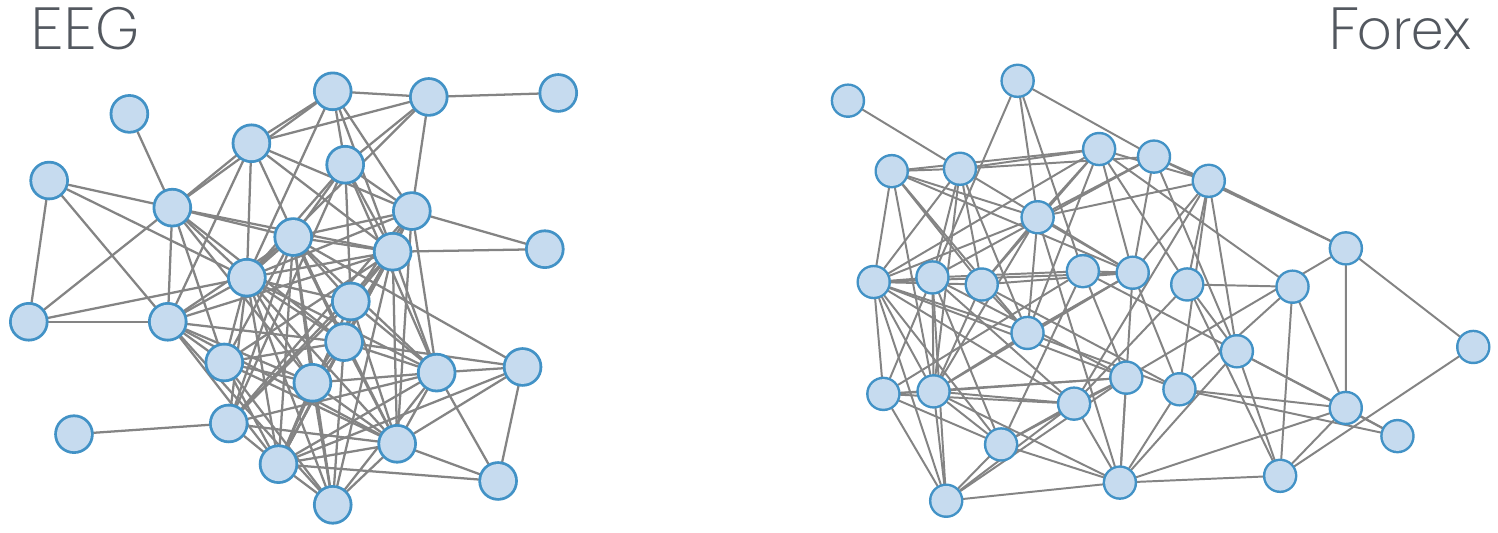}
\caption{ Graphical representation of a brain functional network (left), and one Forex network (right), using the optimal reconstruction parameters for the latter ($l = 30$, $\tau = 0.31$, and $\gamma = 95$).  }
\label{fig:Networks}
\end{figure}

A graphical representation of two networks is presented in Fig. \ref{fig:Networks}; note that the structure of both is qualitatively very similar, with a marked core-periphery configuration.

\newpage
\clearpage

\bibliography{BrainIdent}

\end{document}